\documentclass[11pt, dvipsnames]{article}
\usepackage[margin=1in]{geometry}
\pdfoutput =1
\usepackage{setspace}
\usepackage[round,sort&compress]{natbib}
\let\cite\citep
\usepackage{amsmath}
\usepackage{amssymb}
\usepackage{algorithm}
\usepackage{varwidth}
\usepackage{booktabs}
\usepackage{algpseudocode}
\allowdisplaybreaks
\usepackage{hyperref}
\makeatletter
\newcommand{\printfnsymbol}[1]{%
  \textsuperscript{\@fnsymbol{#1}}%
}
\makeatother
\usepackage{graphicx}
\usepackage{caption}
\usepackage{natbib}
\usepackage{subcaption}
\usepackage{color}
\usepackage{bbm}
\usepackage{float}

\usepackage{amsthm}
\usepackage{dsfont}
\usepackage{mathrsfs}
\usepackage{lmodern}
\usepackage{hyperref}

\usepackage{mathtools}
\usepackage{verbatim}
\usepackage{subcaption}
\usepackage{enumitem}
\usepackage{amsthm}
\usepackage{macros}
\usepackage{todonotes}

\newtheorem{theorem}{Theorem}[section]

\newtheorem{proposition}[theorem]{Proposition}

\newtheorem{remark}[theorem]{Remark}

\newtheorem{definition}[theorem]{Definition}

\usepackage{xcolor}

\DeclareMathOperator*{\argmin}{arg\,min}

\usepackage{appendix}
\title{Congestion Pricing for Efficiency and Equity: Theory and Applications to the San Francisco Bay Area
\thanks{This version: September, 2024. This work is supported by NSF Collaborative Research: Transferable, Hierarchical, Expressive, Optimal, Robust, Interpretable NETworks (THEORINET) under award No. DMS-2031899.}}
\author{Chinmay Maheshwari, Kshitij Kulkarni, Druv Pai, \\  Jiarui Yang, Manxi Wu, Shankar Sastry \thanks{C. Maheshwari (chinmay\_maheshwari@berkeley.edu), K. Kulkarni (kshitijkulkarni@berkeley.edu), D. Pai (druvpai@berkeley.edu), and S. Sastry (shankar\_sastry@berkeley.edu) are with the Department of Electrical Engineering and Computer Sciences at the University of California, Berkeley. M. Wu (manxiwu@cornell.edu) is with the School of Operations Research and Information Engineering at Cornell University. J. Yang (jy925@cornell.edu) is with School of Operations Research and Information Engineering at Cornell Tech. }}
\date{}

\begin{document}

\maketitle
\begin{abstract}
     Congestion pricing, while adopted by many cities to alleviate traffic congestion, raises concerns about widening socioeconomic disparities due to its disproportionate impact on low-income travelers. We address this concern by proposing a new class of congestion pricing schemes that not only minimize total travel time, but also incorporate an equity objective, reducing disparities in the relative change in travel costs across populations with different incomes, following the implementation of tolls. Our analysis builds on a congestion game model with heterogeneous traveler populations. We present four pricing schemes that account for practical considerations, such as the ability to charge differentiated tolls to various traveler populations and the option to toll all or only a subset of edges in the network. We evaluate our pricing schemes in the calibrated freeway network of the San Francisco Bay Area. We demonstrate that the proposed congestion pricing schemes improve both the total travel time and the equity objective compared to the current pricing scheme. 
 Our results further show that pricing schemes charging differentiated prices to traveler populations with varying \votTag\ lead to a more equitable distribution of travel costs compared to those that charge a homogeneous price to all.
\end{abstract}
\newpage

\newpage

\section{Introduction}
Congestion
pricing is an incentivize mechanism for effective utilization of road infrastructure among selfish travelers. Widely adopted in many major cities, both theoretical \citep{yang2005mathematical} and empirical \citep{craik2023equity, phang2004road, percoco2015heterogeneity, eliasson2006equity} studies have shown that congestion pricing can reduce traffic congestion and greenhouse gas emissions, and improve air quality \citep{liu2020commuters, lebeau2019traffic,zhang2013air,han2022effect}. The revenue generated from congestion pricing is often reinvested to improve the road infrastructure, public transit, and other sustainable mobility initiatives \citep{goodwin1990make, small1992using}. Despite these benefits, implementation of congestion pricing often faces challenges, and one of the primary concerns is its disproportional impact on low-income travelers \citep{primerincome, gemici2018wealth}. 
These travelers often have limited access to alternative transportation options, and the additional financial burden of congestion fees may exacerbate existing inequalities.

{In this work we present a principled approach to compute congestion pricing schemes that incorporate both \textit{(i)} the \emph{efficiency} objective of minimizing the total travel time on the network, and \textit{(ii)} the \emph{equity-welfare} objective, where the equity is assessed in terms of maximum disparity in relative change in travel costs experienced by different traveler populations following the implementation of tolls, compared to a scenario with no tolls, and welfare is assessed as the average relative change in travel costs experienced by travelers across all types following the implementation tolls, compared to scenario with no tolls.}

We consider a non-atomic routing game, where travelers make routing decisions based on the travel time of each route plus the monetary cost that includes tolls and gas prices. The monetary cost is adjusted by the travelers' \votTag --- the amount of money a traveler is willing to pay to save a unit of time. Our game has a finite number of traveler populations, each with a heterogeneous \votTag. Following the result from \cite{guo2010pareto}, the equilibrium flow in our game is unique, and can be computed by solving a convex optimization problem. Moreover, the congestion minimizing edge flow vector (i.e. the edge flow vector that minimizes the total travel time) is unique.

We propose four kinds of congestion pricing schemes that differ in terms of whether (a)  tolls are differentiated based on the type of population, and (b) tolls can be set on all edges or a subset of edges. 
In particular, the four congestion pricing schemes are: 
(i) \emph{homogeneous pricing scheme with no support constraints}, denoted by \homo, where all populations are charged with the same tolls and all edges are allowed to be tolled; 
(ii)  \emph{heterogeneous pricing scheme with no support constraints}, denoted by \het, where populations are charged with differentiated 
toll prices based on their types and all edges can be tolled;  
(iii) \emph{homogeneous pricing scheme with support constraints}, denoted by \homosc, where tolls are not differentiated but only a subset of edges can be tolled; 
(iv) \emph{heterogeneous pricing scheme with support constraints}, denoted by \hetsc, where tolls are differentiated and only a subset of edges are tolled.

We compute the tolls in each pricing scheme using a two-step approach. First, we characterize the set of tolls that minimize the total travel time (i.e. {efficiency objective}). 
Second, we select a particular toll price in the set of tolls computed in the first step to optimize for an objective that achieves the trade-off between average welfare of all populations and the equity across different populations. 
Under the \homo\ and \het\ pricing schemes, the set of tolls that minimize the total travel time (as computed in the first step) can be characterized as the set of solutions of a linear program, and the second step of selecting a particular toll price is also an optimal solution of a linear program (Proposition \ref{prop: HomTollsEquitable}) and \het\ (Proposition \ref{prop: HetTollsEquitable}).
The two step approach and the linear program formulations build on the study of enforceable equilibrium flows in routing games with heterogeneous populations \citep{fleischer2004tolls, karakostas2004edge, yang2005mathematical, harks2023unified}. On the other hand, under $\homosc$ and $\hetsc$, direct extensions of the two linear programs to include toll support set constraints are not guaranteed to achieve the efficiency goal. 
In fact, the problem of designing congestion minimizing pricing schemes with support constraints is known to be NP hard without the consideration of heterogeneous \votTag\  \citep{hoefer2008taxing, bonifaci2011efficiency, harks2015computing}. Building on the linear programming based approaches developed for the pricing schemes without support constraints, we propose a linear programming based heuristic to compute tolls with support constraints and evaluate their efficiency outcomes in the case study.

\begin{figure} 
     \centering
         \centering
\includegraphics[width=0.45\textwidth]{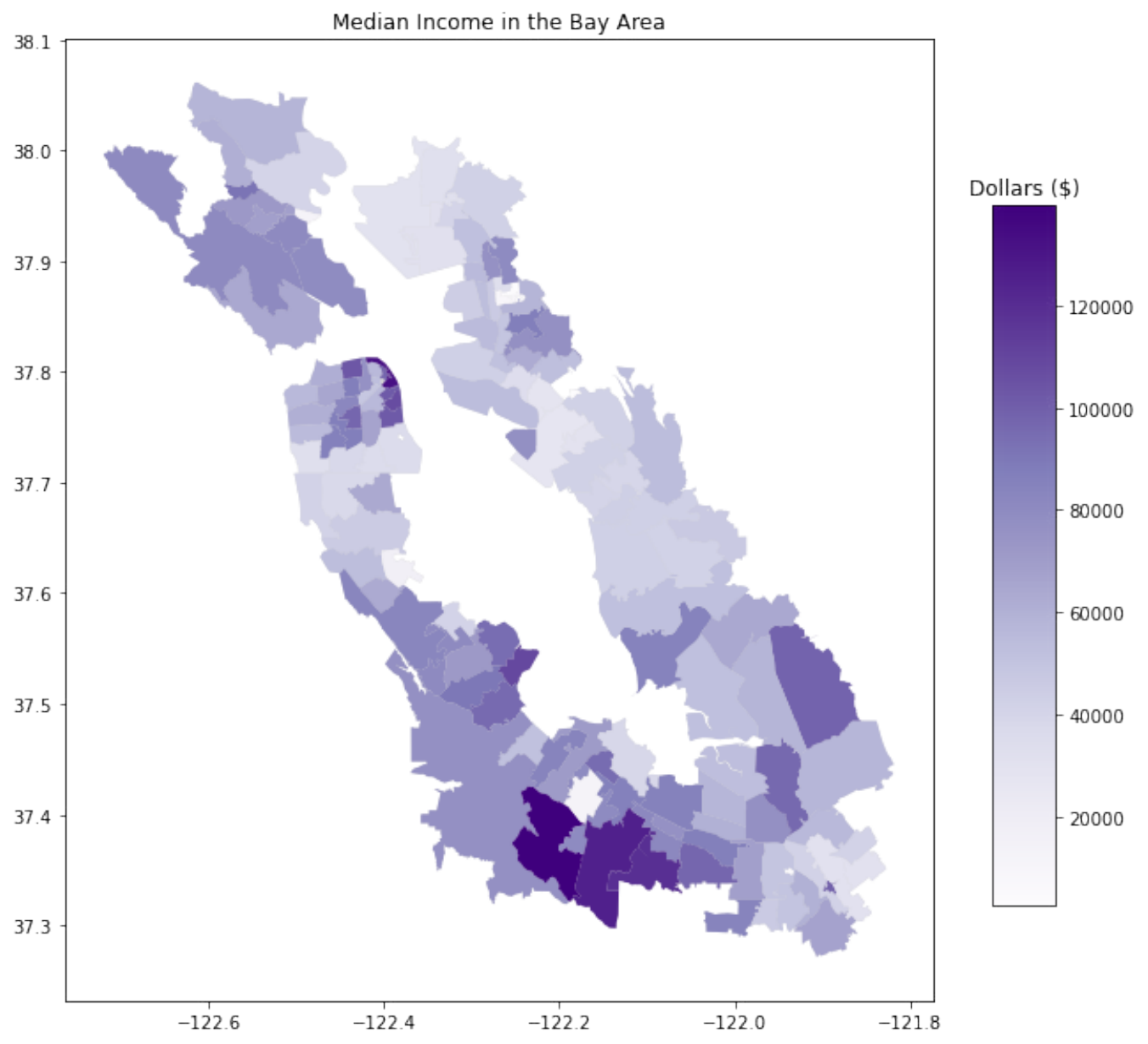}
         \caption{Median income.}
         \label{fig: MedianIncome}
     \end{figure}

\begin{figure}    
         \centering
\includegraphics[width=0.45\textwidth]{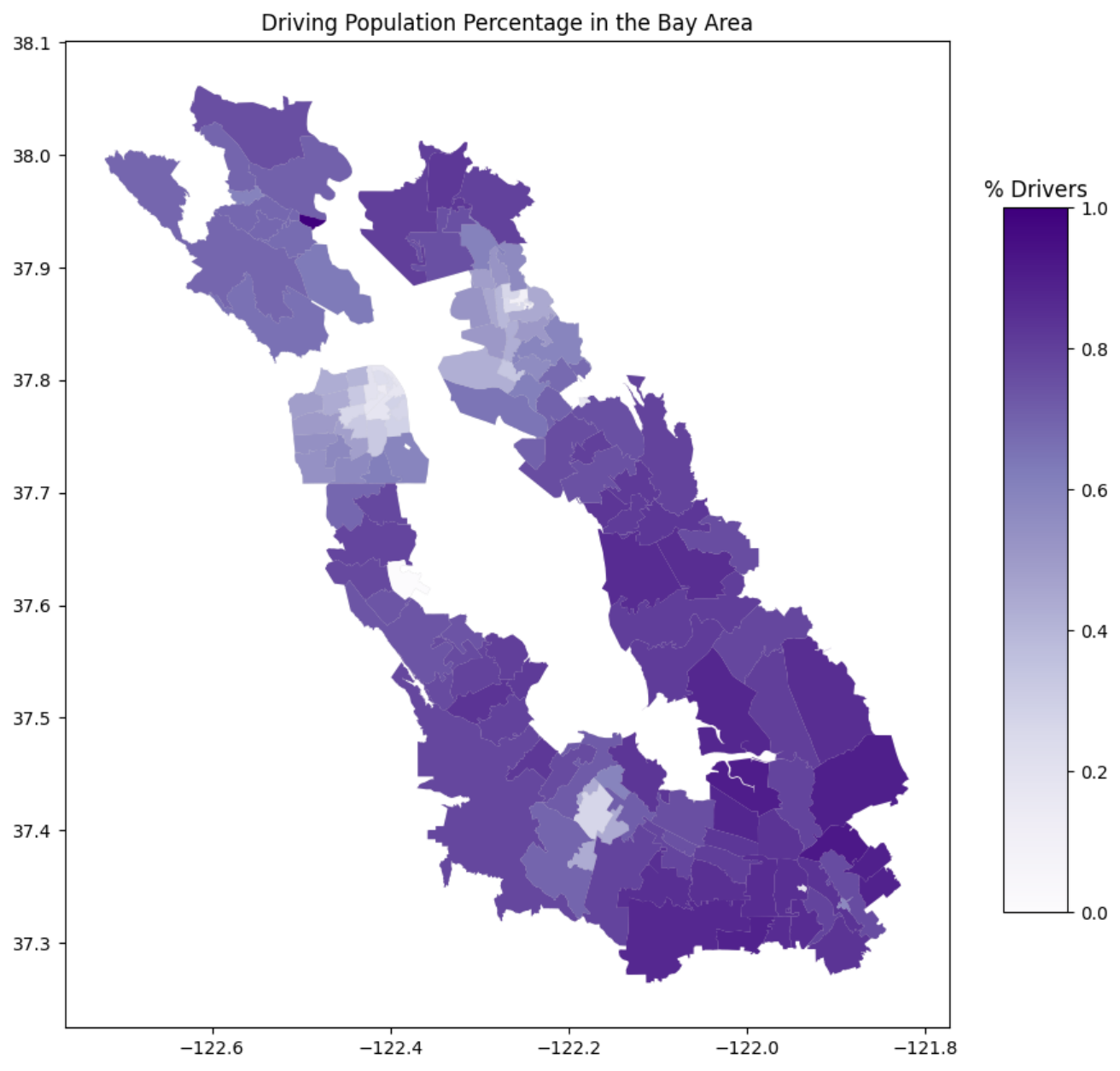}
         \caption{Driving population percentage.}
          \label{fig: DrivingPopulation}
     \end{figure}

We next apply our results to evaluate the performances of the four congestion pricing schemes in the San Francisco Bay Area freeway network.  
Populations in the San Francisco Bay area exhibit significant socioeconomic disparities. This is evident from the distribution of median annual individual income of each neighborhood as shown in Figure \ref{fig: MedianIncome}. Moreover, the area has low public transport coverage and thus majority of the populations commute via car. We can see in Figure \ref{fig: DrivingPopulation} that the driving population percentage of most zip codes outside of San Francisco and Oakland cities are higher than 60\%. Moreover, zip codes that are on the east side of the Bay Area have both a high percentage of driving population and a low median individual income. This observation underscores the importance to design efficient and equitable congestion pricing schemes that account for the socioeconomic and geographic disparities, and the disproportionate impact of tolling on different populations.  


We model the freeway network in the San Francisco Bay Area as a network with 17 nodes (Figure \ref{fig:node_map}). Each node represents a major work or home location for travelers, and the edges represent the primary freeways connecting these locations. Since we differentiate populations based on their \votTag, which is a latent parameter that cannot be directly estimated from the data, we use the median individual income as a proxy (\citep{athira2016estimation, gunn2001spatial, waters1994value, thomas1970value}) to categorize travelers with home at each node into three types of populations with low, middle and high \votTag, respectively.







Using high-fidelity datasets from Safegraph, the Caltrans Performance Measurement System (PeMS), and the American Community Survey (ACS), we calibrate the latency function of each edge and the demand of each traveler population between each pair of nodes.

The current congestion pricing scheme, denoted as \curr\, sets \(\$7\) price on each of the bridges in the Bay Area (Figure \ref{fig:node_map}). We compute the four congestion pricing schemes (\homo, \het, \homosc, \hetsc), and compare the resulting equilibrium routing behavior in comparison to \curr\ and the \zero\ pricing scheme that set no tolls. We summarize our finding below:

\textit{(i) Efficiency and Equity:} All four proposed pricing schemes leads to a lower value of total travel time compared to \curr. Surprisingly, \curr\ is also marginally outperformed by \zero. 
    This is primarily attributed to the fact that the homogeneous toll price of \$7 on all bridges under \curr\ does not account for the heterogeneous distribution of populations between different home-work locations. 
   We show that \homo\ and \het\ achieve the minimum congestion, as indicated by our theoretical result (Proposition \ref{prop: HomHetDualOnly}). Additionally, \homosc\ and \hetsc\ achieve lower value of total travel time than \curr\ and \zero\ but higher than \homo\ and \het. Furthermore, we find that the price of anarchy (POA) -- the ratio between the total travel time in equilibrium with no tolls and that of the minimum value total travel time  \cite{roughgarden2010algorithmic} -- in our setup is 1.04, which is close to 1. This is likely due to high total demand of travelers in the Bay area network since POA always converges to 1 in routing games as the population demand increases \citep{colini2020selfish, cominetti2021price}. 
   
We find that all pricing schemes, except \homo, result in lower travel costs for all traveler populations compared to \curr. Additionally, our results show that \curr\ is outperformed by all other schemes, except \homo, even on the equity metric.

\textit{(ii) Revenue Generation:} We observe that the revenue generated by \homo\ is the highest as it charges high tolls to all travelers in order to achieve the minimum congestion. Moreover, the revenues generated by \het, \homosc\ and \hetsc\ are comparable to \curr\, with \het\ being marginally higher and, \homosc\ and \hetsc, marginally lower.

The rest of the paper is organized as follows: Section \ref{sec: RelatedWorks} presents an overview of past works related to our paper. Section \ref{sec: Model} presents the model of routing games with heterogeneous populations. {Section \ref{sec:method}} presents the computation methods for the four congestion pricing schemes (\homo, \het, \homosc, and \hetsc). Section \ref{sec: ModelCalibrationBayArea} presents calibration of the routing game model in the San Francisco Bay Area. Section \ref{sec: Analysis} presents the efficiency and equity evaluation of the proposed pricing schemes and the comparison of the emerging congestion patterns. 

\section{Related works}\label{sec: RelatedWorks}
The literature on designing congestion pricing schemes can be categorized into two main threads: \emph{first-best} and \emph{second-best}. {First-best} pricing schemes allow tolls to be placed on every edge of the network. The most popular first-best tolling scheme is marginal cost pricing, which sets the toll price to be the marginal cost created by an additional unit of congestion on each edge.
\cite{arnott1994economics, beckmann1956studies, smith1979marginal,roughgarden2010algorithmic}. 
Additionally, an extensive line of research in this thread also focuses on characterizing the set of all congestion-minimizing toll prices (see \cite{yang2005mathematical}, and references therein).
On the other hand, {second-best} pricing schemes restrict the set of edges that can be tolled. The literature on second-best pricing schemes primarily focuses on formulating the problem as a mathematical program with equilibrium constraints (MPEC) and developing algorithms to approximate the optimal solution (e.g. \citep{yang1996optimal, brotcorne2001bilevel, ferrari2002road, labbe1998bilevel, larsson1998side, lim2002transportation, patriksson2002mathematical, verhoef2002second, lawphongpanich2004mpec, ekstrom2009heuristic, kalashnikov2016heuristic}). 
The papers \cite{hoefer2008taxing, bonifaci2011efficiency,harks2015computing} studied the problem of characterizing the hardness of the problem of designing second-best tolls. 
The paper \cite{hoefer2008taxing} showed that it is NP hard to compute optimal tolls on a subset of edges in general networks and gave a polynomial time algorithm to solve the problem for the parallel link case with affine latency functions. This was extended to allow for non-affine latency functions by \cite{harks2015computing}, and upper bound on the toll values in
\cite{bonifaci2011efficiency}. 
In our setup, \homo \ and \het \ are first-best pricing schemes and \homosc \ and \hetsc \ are second-best pricing schemes. We contribute to this line of literature by proposing a multi-step linear programming based approach to compute \homo \ and \het \ that account for the equity objective and the heterogeneous traveler populations. Our approach is also an efficient heuristic to solve \homosc \ and \hetsc \ with atmost three linear programs instead of iteratively computing the Wardrop equilibrium.  

The literature on congestion pricing has mostly focused on homogeneous pricing schemes with a few exceptions. 
The paper \cite{feng2023collaborative} considered tolling schemes that  differentiate conventional vehicles from clean energy vehicles. Moreover, differentiated tolls are also used in \citep{lazar2021role, lazar2020optimal, mehr2019pricing} to study mixed autonomy. 
The paper \cite{brown2016study} studied the impact of differentiated tolling in parallel-link networks with affine cost functions and travelers that have heterogeneous \votTag.

One effort to ameliorate the inequities resulting from congestion pricing is to redistribute the toll revenue 
(see \cite{small1992using, goodwin1989rule, daganzo1995pareto, adler2001direct, guo2010pareto, jalota2021efficiency}, or references therein), 
or provide tradable or untradable travel credits (see  \cite{zhu2015properties, lin2021credit, dogterom2017tradable, nie2012transaction,wu2012design}, or references therein). 
The papers \cite{goodwin1990make, small1992using} were amongst the first to propose different ways to redistribute the revenue in form of infrastructure development and tax rebates. The effectiveness of redistribution schemes are theoretically analyzed in single-lane bottleneck models \cite{arnott1994economics, bernstein1993congestion}, parallel networks \cite{adler2001direct}, and single origin-destination network \cite{eliasson2001road}.

\emph{Pareto-improving} congestion pricing schemes were introduced as another approach to reduce inequality. First proposed by \cite{lawphongpanich2007pareto}, Pareto-improving congestion pricing minimizes the total congestion while ensuring that no travelers are worse off in comparison to no tolls. The paper \cite{song2009nonnegative} studied the design of Pareto-improving schemes for travelers with heterogeneous \votTag, and \citep{lawphongpanich2010solving} further proved that such Pareto-improving schemes only exist in special classes of networks. The paper \cite{guo2010pareto} studied the problem of designing Pareto-improving pricing schemes combined with revenue refund. \cite{jalota2021efficiency} extended this line of research by developing optimal revenue refunding schemes to minimize the congestion and inequity together. In both \cite{guo2010pareto} and \cite{jalota2021efficiency}, the tolls minimize the weighted sum of travel times with weights being each population's \votTag. This objective is different from our goal of minimizing the unweighted total travel time, which is a more suitable metric to assess the environmental impact of congestion.

The third approach to addressing inequality is the study of \emph{fairness constrained traffic assignment problem} proposed by \cite{jahn2005system}, where the fairness metric is the maximum difference of travel time experienced by travelers between the same origin-destination pair. \cite{angelelli2016proactive, angelelli2021system} extended this line of research by developing algorithmic methods to solve the fairness constrained traffic assignment problem. The problem of devising congestion pricing schemes which could enforce the resulting traffic assignment patters was studied in \cite{jalota2023balancing}. Particularly, \cite{jalota2023balancing} studies homogeneous pricing scheme that implements the traffic assignment minimizing an interpolation of the potential function (which is used to characterize the equilibrium) and the social cost function. 

Our work contributes to all of the above studies on the equity of congestion pricing from three aspects: 
\textit{(i)} Our equity consideration accounts for both the travel time cost and the monetary cost that includes both the toll and the gas prices. This generalizes the fairness notion that focuses only on the travel time difference; 
\textit{(ii)} Our tolling scheme minimizes the total congestion in the network (i.e. guarantees the optimal efficiency) while providing the central planner a flexible way to trade-off between the total welfare, equity across heterogeneous populations and total revenue. In particular, by tuning the parameter that governs the trade-off between the average welfare and equity, we can increase or reduce the revenue collected by the our pricing scheme; 
\textit{(iii)} We provide a comprehensive evaluation of different congestion pricing schemes in terms of efficiency, equity and revenue using real-world data collected in the San Francisco Bay Area. 

Another line of research related to this paper is on developing inverse optimization based tools to estimate model parameters in non-atomic routing games such as demand, latency functions etc \cite{wollenstein2019joint, bertsimas2015data, zhang2016price}. There are several differences between our approach and these works.  First, we use  high fidelity datasets to directly estimate the latency on every edge and the demand of travelers. Second, we consider heterogeneous population of travelers as opposed to the homogeneous population of travelers considered in these works.  

Finally, on the empirical side, \cite{frick1996bay, nakamura2002congestion, barnes2012impact,  zhang2011congestion} focused on understanding the impact of congestion pricing of the San Francisco-Oakland Bay Bridge, which is the most heavily congested segment in the San Francisco Bay Area highway network.
Our work generalizes this line of work to the entire Bay Area highway network using high-fidelity mobility and socioeconomic datasets.

\section{Model}\label{sec: Model}
In this section, we introduce the non-atomic networked routing game model that forms the basis for our theoretical and computational results. We introduce equilibrium routing and the four types of congestion pricing schemes we consider in this paper. 

\subsection{Network}\label{subsec:network}
Consider a transportation network \(\graph =(\nodes,\edges)\), where \(\nodes\) is the set of nodes, and  \(\edges\) is the set of edges. A set of non-atomic travelers (agents) make routing decisions in the network between their origin and destination. We denote the set of origin-destination (o-d) pairs as \(\odPairs\) and the set of routes (i.e. sequences of edges) connecting each o-d pair \(\od\in \odPairs\) as \(\paths^{\od}\).

Travelers for each o-d pair $k$ are grouped into \(I\) populations, where each population is associated with a different level of \emph{\votTag} \(\vot^{\type} \in \mathbb{R}_{\geq 0}\) that captures the trade-off travelers in population $i$ are willing to make between travel time and monetary cost while selecting between different routes. 
We refer to agents with \votTag\ $\theta^i$ as type $i$ agents.
The demand vector is given by \(\demand = (\demand^{\type\od})_{i\in I, k\in K}\), where \(\demand^{ik}\) is the demand of agents with type $i$ that want to travel between o-d pair $\od$. Throughout this paper, we operate under the \emph{inelastic demand} assumption: traveler demands on each origin-destination pair are constant. This assumption is reasonable given that \textit{(i)} our analysis focuses on the commuting behavior during the morning rush hour, when the majority of trips are work-related with little elasticity; \textit{(ii)} the availability of public transit is sparse and the cost of car ownership is high \citep{Depillis_Lieberman_Chapman_2023}.

The strategy distribution of agents is denoted $\pathFlow = (\pathFlow^{\type\od}_{r})_{r \in R^k, \type \in I, \od \in K}$ , where $\pathFlow^{\type\od}_{r}$ is the flow of agents with type $i$ and o-d pair $\od$ who take route $r$. Therefore, given a demand \(D\), the set of feasible strategy distributions is given by: 

{\small \begin{align}\label{eq: SetFeasible} 
    \mathcal{Q}(D) \doteq \Bigg\{q: \sum_{\route \in \paths^{\od}} \pathFlow_{\route}^{\type\od} = \demand^{\type\od}, \pathFlow_{\route}^{\type\od} \geq 0, \forall \route \in \paths^{\od},  \type \in \votGroup, \od \in \odPairs\Bigg\}.
\end{align}}
Given a strategy distribution \(\pathFlow\in \mathcal{Q}(D)\), the flow of agents of type \(i\in I\) on edge \(\edge\in \edges\) is given by 
\begin{align}\label{eq:wei}
    \popuFlow_{\edge}^{\type}(\pathFlow) \doteq \sum_{\od \in\odPairs}\sum_{\route \in \paths^{\od}} \pathFlow_{\route}^{\type\od} \mathbbm{1}(\edge\in \route),
\end{align} 
and the total flow of agents on edge $\edge \in \edges$ is
\begin{equation}\label{eq: EdgeFlows}
    \edgeFlow_{\edge}(\pathFlow) \doteq \sum_{\type \in \votGroup} \popuFlow_{\edge}^{\type}(\pathFlow).
\end{equation}
The travel time experienced by agents taking edge \(\edge \in \edges\) is \(\latency_{\edge}(\edgeFlow_{\edge}(\pathFlow))\), where the latency function \(\latency_{\edge} \colon \mathbb{R}_+ \to \mathbb{R}_+\) is \emph{continuous, strictly increasing, and convex.} 
Consequently, the total travel time  experienced by agents from o-d pair \(\od \in \odPairs\) who use route \(\route \in \paths^{\od}\) is given by \(\latencyRoute_{\route}(\pathFlow) \doteq \sum_{\edge \in \route}\latency_{\edge}(\edgeFlow_{\edge}(\pathFlow)).\)
With slight abuse of notation, we use \(\ell_r(q)\) and \(\ell_r(w)\) interchangeably to represent the latency of route $r$ where \(w\) is the edge flow vector corresponding to the strategy distribution \(q\). In addition to the travel time, the total cost experienced by each individual agent also includes the congestion price imposed by the planner, and the gas cost required to travel on the route the agent chooses. In particular, let \(\tolls_e^{i}\) be the toll price imposed on travelers of type \(i\in\votGroup\) for using edge \(e\in\edges\), and \(\gasPrice_e\) be the gas cost of using an edge \(e\in \edges\). Note that we allow for the toll price to be type-specific in the general setting. We will later discuss different scenarios for setting the toll prices.  
Given the tolls \(\tolls= (\tolls_{\edge}^{\type})_{\edge\in\edges,\type\in\votGroup}\), the cost experienced by travelers of type \(\type \in \votGroup\) associate with o-d pair \(\od \in \odPairs\) and taking route \(\route \in \paths^{\od}\) is given by 
\begin{align}\label{eq: TotalCost}
\edgeCost_{\route}^{\type}(\pathFlow, \tolls) \doteq \latencyRoute_{\route}(\pathFlow) + \frac{1}{\vot^{\type}}\sum_{\edge \in \route}(\tolls_{\edge}^{\type}+\gasPrice_e). 
\end{align}
Crucially, a key feature of our model is that the toll and gas costs experienced by each agent are modulated by the \votTag\ $\vot^\type$ of that agent. This allows us to model the heterogeneity present in the types of travelers. Given this setup, we define Nash equilibrium to be the strategy distribution such that no traveler has incentive to deviate from their chosen route. That is, 
\begin{definition}\label{def: NashEq} 
Given tolls \(\tolls\), a strategy profile \(\pathFlow^{*}(\tolls)
\) is a Nash equilibrium if 
\begin{align*}
&\forall \type \in \votGroup, \od\in\odPairs, \route\in\paths^{\od}, \quad \pathFlow_{\route}^{\type\od*}(\tolls) > 0 \quad \\ &\Rightarrow \quad \edgeCost_{\route}^{\type}(\pathFlow^{*}(\tolls), \tolls) \leq \edgeCost_{r'}^{\type}(\pathFlow^{*}(\tolls), \tolls)\ \quad \forall r' \in \paths^{\od}.
\end{align*}
\end{definition}

The objective of the planner is to minimize the network congestion, measured by the total travel time experienced by all travelers. For any strategy distribution $q$, we denote the planner's cost function as follows: 
\begin{align}
    S(q) \doteq \quad & \sum_{\edge \in \edges}\edgeFlow_{\edge}(q)\latency_{\edge}(\edgeFlow_{\edge}(q))\label{eq: SocOptProblem}, 
\end{align}
where \(w_{e}(q)\) is given by \eqref{eq: EdgeFlows}. We denote the set of socially optimal strategy distributions as \(q^\dagger \doteq \argmin_{q\in \mathcal{Q}(D)} S(q)\), and the induced socially optimal edge flows as $w^{\dagger}= (w^{\dagger}_e)_{e \in E}$, where $w^{\dagger}_e= w_e(q^{\dagger})$ given by \eqref{eq: EdgeFlows}.
\subsection{Congestion pricing}\label{ssec: TollPrices}
We now introduce two practical considerations for toll implementation. The first consideration is whether or not the toll is type-specific. In particular, a congestion pricing scheme is \emph{homogeneous} if the toll is uniform across all population types, and \emph{heterogeneous} if the toll varies with population types (formally, whether $\tolls_\edge^{\type}$ is allowed to depend on $\type$ or not, on each edge). {The challenge of implementing a heterogeneous scheme is that the population type (i.e. \votTag) is a latent variable that is privately known only by the individual traveler.} In practice, an individual's \votTag\ is often closely correlated with their income level, i.e. higher-income groups are typically associated with a higher \votTag, while lower-income groups correlate with a lower \votTag\ \citep{athira2016estimation, gunn2001spatial, waters1994value, thomas1970value}. Therefore, one way to implement heterogeneous tolling is to set tolls based on the income level of travelers. For example, low income groups, which have significant overlap with the population of low \votTag\ travelers, may receive a subsidy or a toll rebate in certain areas. Such toll relief programs have been established in several states in the United States, e.g. California 
\footnote{\url{https://mtc.ca.gov/news/new-year-brings-new-toll-payment-assistance-programs}}
, Virginia 
\footnote{\url{https://www.vdottollrelief.com/}}
, New York 
\footnote{\url{https://new.mta.info/fares-and-tolls/bridges-and-tunnels/resident-programs}} 
etc.

The second consideration is whether or not tolls can be set on all the edges of the network or only on a subset (formally, whether or not $\tolls_{\edge}^{\type}$ is allowed to be strictly positive on all $\edge \in \edges$). In practical terms, congestion pricing often requires the installation of toll collection facilities, which might not be feasible on all road segments. Thus, a congestion pricing scheme has no support constraints if tolls can be imposed on all edges, or has support constraints if tolls can only be imposed on a subset of edges, denoted as \(\edges_T\). We note that congestion pricing schemes with (resp. without) support constraints are also referred as first-best (resp. second-best) tolling schemes in literature. 

Building on the above two considerations, we define four types of tolling schemes: \textit{(i)} \emph{Homogeneous tolls with no support constraints} ($\homo$): \(p_e^i\geq 0\) and \(p_e^i=p_e^j\) for all \(e\in \edges\) and all \(i,j\in I\); \textit{(ii)} \emph{Heterogeneous tolls with no support constraints} ($\het$): \(p_e^i\geq 0\) for all \(e\in \edges, i\in I\); \emph{Homogeneous tolls with support constraints} ($\homosc$): \(p_e^i=p_e^j\) for all $i, j \in I$ and all $e \in E$. Additionally, \(p_e^i = 0\) for all \(e\in \edges\backslash\edges_T \), and \(p_e^i \geq 0\) for all \(e\in \edges_T\).\textit{(iii)}
Building on the above two considerations, we define four types of tolling schemes: \textit{(i)} \emph{Homogeneous tolls with no support constraints} ($\homo$): \(p_e^i\geq 0\) and \(p_e^i=p_e^j\) for all \(e\in \edges\) and all \(i,j\in I\); \textit{(ii)} \emph{Heterogeneous tolls with no support constraints} ($\het$): \(p_e^i\geq 0\) for all \(e\in \edges, i\in I\); \textit{(iii)} \emph{Homogeneous tolls with support constraints} ($\homosc$): \(p_e^i=p_e^j\) for all $i, j \in I$ and all $e \in E$. Additionally, \(p_e^i = 0\) for all \(e\in \edges\backslash\edges_T \), and \(p_e^i \geq 0\) for all \(e\in \edges_T\); \textit{(iv)} \emph{Heterogeneous tolls with support constraints} ($\hetsc$): \(p_e^i = 0\) for all \(e\in \edges\backslash\edges_T \), and \(p_e^i \geq 0\) for all \(e\in \edges_T\).

\section{Computation methods}\label{sec:method} 
In this section, we outline methods for computing equilibrium routing strategies and the four congestion pricing schemes.
We first establish that, given any fixed toll values, the equilibrium outcome can be derived as the optimal solution to a convex optimization problem.
We then demonstrate that the set of homogeneous tolls ($\homo$) and heterogeneous tolls ($\het$) without support constraints that realize the socially optimal edge flows can be characterized as the set of optimal solutions of linear programs.
Next, we present a {multi-step} approach for calculating the toll prices that strikes a balance between equity, as measured by the cost disparity between travelers from different populations, and at the same time, maximizing the welfare of all traveler populations.
For congestion pricing schemes with support constraints, we adapt our approach to provide a heuristic for calculating $\homosc$ and $\hetsc$, acknowledging that such solutions may not guarantee the implementation of the socially optimal edge flows.

\begin{proposition}\label{prop: PotentialFunction}Given toll prices \(\tolls\), a strategy distribution \(q^\ast(\tolls)\) is a Nash equilibrium if and only if it is a solution to the following convex optimization problem:
\begin{align}
    \min_{\pathFlow \in \mathcal{Q}(D)} \quad  \Phi(\pathFlow, \tolls, \vot) & = \sum_{\edge \in \edges} \int_{0}^{\edgeFlow_{\edge}(\pathFlow)} \latency_{\edge}(z)\ \mathrm{d}z \notag \\
    &\quad + \sum_{\type \in \votGroup}\sum_{\edge \in \edges} \frac{(\tolls_{\edge}^{\type}+\gasPrice_e)}{\vot^{\type}} \popuFlow_{\edge}^{\type}(\pathFlow), \label{eq: PotentialCongestionGame} 
\end{align}
where \(w_{e}^{i}(q)\), \(w_{e}(q)\) are given by \eqref{eq:wei} and \eqref{eq: EdgeFlows}, respectively.   Moreover, given any toll price vector $p$, the equilibrium edge flow vector \(w^*(\tolls):= w(q^*(\tolls))\) is unique. Additionally, the socially optimal edge flow vector \(w^{\dagger}\) is unique. 
\end{proposition}

\cite{guo2010pareto} showed the same result as Proposition \ref{prop: PotentialFunction} without the gas price. 
The proof follows directly from \cite{guo2010pareto}, and is available in the online extended version of this article \cite{maheshwari2024congestion}.

Proposition \ref{prop: PotentialFunction} shows that $\edgeFlow^{\dagger {}}$ is unique. However, we note that such a $\edgeFlow^{\dagger{}}$ may be induced by multiple type-specific flow vectors $\popuFlow^{\dagger{}}$. Although these different type-specific flow vectors all induce the same aggregate edge load, and thus minimize the total cost, they may lead to different travel times experienced by different populations. 

The following proposition shows that the set of prices $\homo$ (resp. $\het$) that implements the socially optimal edge load can be characterized each by a linear program. 

\begin{proposition}\label{prop: HomHetDualOnly}
    \begin{itemize}
        \item[(1)] 
       A homogeneous congestion pricing scheme \(p^\dagger = (p^\dagger_e)_{e\in E}\) implements the socially optimal edge flow \(w^{\dagger{}} = (w^{\dagger{}}_e)_{e\in E}\) if and only if there exists \(z^\dagger\) such that $(p^\dagger, z^\dagger)$ is a solution to the following linear program:
    \begin{equation}\label{eq: HomToll}\tag{\(\mathcal{P}_{\textsf{hom}}\)}
        \begin{aligned}
   T^\ast_{\textsf{hom}} = \max_{\tolls,z} &~ \sum_{i\in I}\sum_{k\in K}D^{ik} z^{ik} - \sum_{e\in \edges}\tolls_ew_e^{\dagger{}},\\ \text{s.t.} \quad &~  z^{ik} - \sum_{e\in r}(\tolls_e+\gasPrice_e) \leq \theta^i\ell_r(w^{\dagger{}}), \\ & \quad \quad \forall k\in K, r\in \paths^k, i\in I, \\
 {} &~ \tolls_e \geq 0, \quad \forall e\in E.
\end{aligned}
\end{equation}
\item[(2)] A heterogeneous congestion pricing scheme \(p^\dagger = (p^{i \dagger}_{e})_{e\in E, i\in I}\) implements a type-specific socially optimal edge flow \(\popuFlow^{\dagger, {}}= (\popuFlow_e^{\dagger i, {}})_{e\in E, i\in I}\) if and only if there exists a \(z^\dagger\) such that \((p^\dagger, z^\dagger)\) is a solution to the following linear program:
\begin{equation}\label{eq: Het}\tag{\(\mathcal{P}_{\textsf{het}}\)}
    \begin{aligned}
    T^\ast_{\textsf{het}}(\popuFlow^{\dagger{}}) = \max_{p,z} & ~ \sum_{i\in I}\sum_{k\in K}D^{ik} z^{ik} - \sum_{e\in E}\sum_{i\in I}p_e^i\popuFlow_e^{\dagger i,{}},\\ 
   \text{s.t.} & ~ z^{ik} - \sum_{e\in r}(p_e^i + \gasPrice_e )\leq \theta^i\ell_r(w^{\dagger{}}), \\&\quad \quad \forall k\in K, r\in \paths^k, i\in I, \\
   & ~ p_e^i \geq 0, \quad \forall e\in E, i\in I.
    \end{aligned}
\end{equation}
    \end{itemize}
\end{proposition}
Proposition \ref{prop: HomHetDualOnly} follows the results in \cite{fleischer2004tolls, karakostas2004edge, marcotte2009existence, yang2005mathematical, harks2023unified}. 
The proof builds on the two linear programs 
\eqref{eq: HomToll} -- \eqref{eq: Het} and their dual 
programs \eqref{eq: HomDual} and \eqref{eq: HetDual} as follows: 
\begin{subequations}
\begin{align}
       \min_{q} &~~ \sum_{i\in I}\sum_{k\in K}\sum_{r\in \paths^k}(\theta^i \ell_r(w^{\dagger{}}) +\sum_{e\in r}g_e)q_r^{ik} \label{eq: HomDual}\tag{\(\mathcal{D}_{\textsf{hom}}\)}
       \\ \text{s.t.} &~~ \sum_{i\in I}\sum_{k\in K}\sum_{r\in \paths^k: e\in r}q_r^{ik} \leq  w_e^{\dagger{}}, \quad \forall  e\in E,  \label{eq: HomDual_a}\tag{\(\mathcal{D}_{\textsf{hom}.a}\)}\\
    {}&~~   \sum_{r\in \paths^k}q_r^{ik}  = D^{ik},\quad \forall  i\in I, k\in K, \label{eq: HomDual_b}\tag{\(\mathcal{D}_{\textsf{hom}.b}\)}\\ 
    {} & ~~q_r^{ik} \geq 0 \quad \forall  i\in I, k\in K, r\in \paths^k. \label{eq: HomDual_c}\tag{\(\mathcal{D}_{\textsf{hom}.c}\)}\\
    \min_q  & ~~ \sum_{i\in I}\sum_{k\in K}\sum_{r\in \paths^k}(\theta^i \ell_r(w^{\dagger{}}) +\sum_{e\in r}g_e)q_r^{ik} \label{eq: HetDual}\tag{\(\mathcal{D}_{\textsf{het}}\)}\\ \text{s.t.} &~~ \sum_{k\in K}\sum_{r\in \paths^k: e\in r}q_r^{ik} \leq  \popuFlow_e^{\dagger i, {}}, \quad \forall  e\in E, i\in I, \label{subeq: het_1}\tag{\(\mathcal{D}_{\textsf{het}.a}\)}\\
    &~~   \sum_{r\in \paths^k}q_r^{ik}  = D^{ik}, \quad \forall  i\in I, k\in K, \label{subeq: het_2}\tag{\(\mathcal{D}_{\textsf{het}.b}\)}\\ 
    &~~ q_r^{ik} \geq 0, \quad \forall i\in I, k\in K, r\in \paths^k.\label{subeq: het_3}\tag{\(\mathcal{D}_{\textsf{het}.c}\)}
\end{align}
\end{subequations}
Under both $\homo$ and $\het$, the feasibility constraints of the associated primal and dual programs as well as the complementary slackness conditions are equivalent to the equilibrium condition where only routes with the minimum cost are taken by travelers. Moreover, constraints \eqref{eq: HomDual_a} and \eqref{subeq: het_1} must be tight at optimality, indicating that the induced flow vector in equilibrium is indeed \(w^\dagger\), which minimizes total travel time. Therefore, the set of optimal solutions of \eqref{eq: HomToll} and \eqref{eq: Het} are the set of toll vectors that induce \(w^\dagger\) under $\homo$ and $\het$, respectively. 

We denote $\Tollhom$ as the set of socially optimal toll prices for $\homo$, and $\Tollhet(\popuFlow^{\dagger})$ as the set of socially optimal toll price for $\het$ that induces a type-specific socially optimal edge flow $\popuFlow^{\dagger{}}$. Proposition \ref{prop: HomHetDualOnly} demonstrates that both sets can be computed as the optimal solution set of linear programs. We note that the set $\Tollhet(\popuFlow^{\dagger{}})$ depends on which type-specific socially optimal flow $\popuFlow^{\dagger}$ is induced since the objective function \eqref{eq: Het} depends on $\popuFlow^{\dagger}$. 

Furthermore, $\Tollhom$ and $\Tollhet(f^{\dagger{}})$ may not be singleton. This presents an opportunity for the planner to decide which specific toll price from the optimal solution set to implement. While all tolls in $\Tollhom$ and $\Tollhet(f^{\dagger{}})$ achieve the minimum social cost, they do so by impacting travelers differently given their individual origin-destination pair and \votTag. We consider that the central planner aims at solving the following problem: 
{\small \begin{equation}\label{eq: planner_obj}
\begin{split}
    &\min_p ~  \plannerobj(p):= \\ & \underbrace{\max_{i,i'\in I} \bigg|\frac{1}{D^i}\sum_{k\in K}D^{ik}\frac{\cikdag(p)}{\cikdag(0)}-\frac{1}{D^{i'}}\sum_{k'\in K}D^{i'k'}\frac{c^{i'k'\dagger}(p)}{c^{i'k'\dagger}(0)}\bigg|}_{(i)}  \\ &\quad \quad \quad + \lambda \underbrace{\frac{1}{D}\sum_{i\in I}\sum_{k\in K}D^{ik}\frac{\cikdag(p)}{\cikdag(0)}}_{(ii)},\\
     &s.t. \quad \quad  p \in \left\{\begin{array}{ll}
         \Tollhom, & \quad \text{in $\homo$,} \\
         \Tollhet(\popuFlow^{\dagger{}}), & \quad \text{in $\het$ with route flow $\popuFlow^{\dagger{}}$,}
     \end{array}
     \right.
\end{split}
\end{equation}}
where $D^i = \sum_{k \in K} D^{ik}$, $D= \sum_{i \in I} D^i$, $\lambda \geq 0$, 
\begin{align}\label{eq:cikdag}
    \cikdag(p) = \min_{r \in R^k} \left\{\ell_r(w^{\dagger{}})+ \frac{1}{\theta^i} \sum_{e \in r} (p_e+g_e)\right\}
\end{align}
is the equilibrium cost of individuals with o-d pair $k$ and type $i$ given the toll price $p$ and socially optimal edge load $\wdag$, and \(\cikdag(0)\) is the equilibrium cost of individuals with o-d pair \(k\) and type \(i\) given no (or zero) tolls. We emphasize that the cost $\cikdag(p)$ is the minimum cost of choosing a route given the socially optimal load vector $\wdag$, the toll price, and the gas fee. 
This is indeed an equilibrium cost of traveler population with type $i$ and o-d pair $k$ since any $p \in \Tollhom$ or $p \in \Tollhet(\popuFlow^{\dagger})$ guarantees that the equilibrium edge vector is $\wdag$.

The objective function in \eqref{eq: planner_obj} indicates that the central planner selects the toll price that not only minimizes the total travel time but also balances the equity among populations with different \votTag, and the average welfare that accounts for the travel time as well as the toll price and gas fee. In particular, in \eqref{eq: planner_obj} \textit{(i)} reflects an equity objective by assessing the maximum disparity in the relative change in travel costs experienced by different types of travelers following the implementation of tolls, compared to a scenario without tolls, and \textit{(ii)} reflects the an average welfare objective that is the average of the relative change in travel costs experienced by all of the travelers following the implementation of tolls, compared to a scenario without tolls. 
Balancing welfare maximization with cost disparity minimization avoids the potential problem with just minimizing cost disparity: charging excessively high tolls to every type of travelers. 
Moreover, $\lambda \geq 0$ is a parameter that governs the relative weight between the equity objective and the welfare objective. 
 
We denote the socially optimal homogeneous congestion pricing scheme that solves the central planner's problem \eqref{eq: planner_obj} as $\phomopt$. The next proposition shows that we can solve the central planner's problem \eqref{eq: planner_obj} for $\homo$ by another linear program. 

\begin{proposition}\label{prop: HomTollsEquitable}
   For the $\hom$ tolling scheme, $\phomopt$ is an optimal solution of the following linear program: 
   {\small\begin{subequations}\label{eq: HomTollsEquitable}
\begin{align}
   &\min_{p,z, y} && y  + \frac{\lambda}{D} \sum_{i\in I}\sum_{k\in K}D^{ik}\frac{z^{ik}}{\theta^ic^{ik\dagger}(0)} \label{eq: HomTollsEquitable_obj}\tag{\(\mathcal{P}^\ast_{\textsf{hom}}\)}\\ & \text{s.t.} && y\geq \frac{1}{D^i}\sum_{k\in K}D^{ik}\frac{z^{ik}}{\theta^i\cikdag(0)}-\frac{1}{D^{i'}}\sum_{k'\in K}D^{i'k'}\frac{z^{i'k'}}{\theta^{i'}c^{i'k'\dagger}(0)}, \notag\\ &{} && \quad\quad \quad  \forall i, i' \in I, \label{eq: HomTollsEquitable_y_1}\tag{\(\mathcal{P}^\ast_{\textsf{hom}.a}\)}\\
   &{} && \sum_{i\in I}\sum_{k\in K}D^{ik} z^{ik} - \sum_{e\in E}p_ew_e^{\dagger{}}\geq T^\ast_{\textsf{hom}}, \label{eq: HomTollsEquitable_opt}\tag{\(\mathcal{P}^\ast_{\textsf{hom}.b}\)}\\
   &{} &&z^{ik} - \sum_{e\in r}(p_e+\gasPrice_e) \leq {\theta^i}\ell_r(w^{\dagger{}}), \  \forall k\in K, r\in \paths^k, i\in I, \label{eq: HomTollsEquitable_eq}\tag{\(\mathcal{P}^\ast_{\textsf{hom}.c}\)}\\
    &{} && p_e \geq 0 \quad  \forall e\in \edges, \label{eq: HomTollsEquitable_nonneg}\tag{\(\mathcal{P}^\ast_{\textsf{hom}.d}\)}
\end{align}
\end{subequations}}
where $T^\ast_{\textsf{hom}}$ is the optimal value of \eqref{eq: HomToll}. 
\end{proposition}
In \eqref{eq: HomTollsEquitable_obj}, constraints \eqref{eq: HomTollsEquitable_eq} and \eqref{eq: HomTollsEquitable_nonneg} ensure that variables $(p, z)$ are in the feasible set of \eqref{eq: HomToll}, and constraint \eqref{eq: HomTollsEquitable_opt} further restrict that the set of $(p, z)$ in \eqref{eq: HomTollsEquitable_obj} to be the set of optimal solutions of \eqref{eq: HomToll}. {Thus, following Proposition \ref{prop: HomHetDualOnly}, any feasible $p$ in \eqref{eq: HomTollsEquitable_obj} must be a toll vector that induces the socially optimal edge flow $\wdag$. 
Moreover, the proof of Proposition \ref{prop: HomHetDualOnly} further ensures that for every \(i\in I, k\in K\) there exists \(r\in R^k\) such that the corresponding constraint in \eqref{eq: HomTollsEquitable_eq} {must be tight at optimum}, which indicates that any $z^{ik}$ in \eqref{eq: HomTollsEquitable_obj} equals to $\theta ^i \cdot c^{ik\dagger}(p)$. Additionally, constraints \eqref{eq: HomTollsEquitable_y_1} guarantee that at optimality $y = \max_{i,i'\in I} \bigg|\frac{1}{D^i}\sum_{k\in K}D^{ik}\frac{\cikdag(p)}{\cikdag(0)}-\frac{1}{D^{i'}}\sum_{k'\in K}D^{i'k'}\frac{c^{i'k'\dagger}(p)}{c^{i'k'\dagger}(0)}\bigg|$. 
Thus, the linear program \eqref{eq: HomTollsEquitable_obj} computes the homogeneous toll prices that minimize the total travel time and optimize the equity and welfare objectives with a relative weight $\lambda$. }

To summarize, the programs \eqref{eq: HomToll} and \eqref{eq: HomTollsEquitable_obj} provide a \emph{two-step approach} for computing $\phomopt$: first, compute $T^\ast_{\textsf{hom}}$ by solving the linear program \eqref{eq: HomToll} given the unique edge flow $\wdag$. Second, compute $\phomopt$ by solving the linear program \eqref{eq: HomTollsEquitable_obj} using $T^\ast_{\textsf{hom}}$.

Next, we show that the central planner can compute the heterogeneous toll price vector (\het) that minimize the total travel time and optimize the equity-welfare objectives \eqref{eq: planner_obj}, denoted as $\phetopt$, using a similar approach as described above. However, in $\het$, one additional issue arises as the set  $\Tollhet(\popuFlow^{\dagger})$ and consequently $\phetopt$ depend on the selection of the type-specific flow vector $\popuFlow^{\dagger}$, which may not be unique. Here, we propose to select the type-specific flow vector $\popuFlow^{\dagger}$ as the one that induces the edge flow vector $\wdag$ (which minimizes total travel time) while also minimizing the disparity in the total travel time experienced across all traveler populations. To compute such a $\popuFlow^{\dagger}$, we first find a feasible routing strategy profile $q^{\dagger}$ that induces $\wdag$ and minimizes the average cost difference among traveler populations. Such a $q^{\dagger}$ can be solved by the following linear program: 
\begin{equation*}\label{eq: cost_diff_min}
\begin{aligned}
    &\min_{q} && x,  \\
    &\text{s.t.} &&  x \geq \sum_{k\in K}\sum_{r\in R^k}\left(q_r^{ik}\ell_r(w^{\dagger{}}) -q_r^{i'k}\ell_r(w^{\dagger{}})\right), \forall i, i' \in I, \\
    & &&\sum_{r\in R^k} q_r^{ik} = D^{ik},  \forall \ i\in I, k\in K,\\ &{} && \sum_{i\in I}\sum_{k\in K}\sum_{r\in R^k: e \in r}q_r^{ik}= w^{\dagger {}}_e,  \forall e\in E,  \\ 
    &{} && q_r^{ik}\geq 0, \forall \ i\in I, k\in K, r\in R^k.
\end{aligned}
\end{equation*}
Then, the induced population-specific flow vector $\popuFlow^{\dagger}$ associated with $q^{\dagger}$ is given by \eqref{eq:wei}. Based on $\popuFlow^{\dagger}$, we compute $\phetopt$ as the optimal solution of a linear program.

\begin{proposition}\label{prop: HetTollsEquitable}
   For the $\het$ tolling scheme, given $\popuFlow^{\dagger}$, $\phetopt$ is an optimal solution of the following linear program: 
   {\small \begin{subequations}
\begin{align}
   &\min_{p,z, y} && y  + \frac{\lambda}{D} \sum_{i\in I}\sum_{k\in K}D^{ik}\frac{z^{ik}}{\theta^ic^{ik\dagger}(0)} \label{eq: HetTollsEquitable}\tag{\(\mathcal{P}^\ast_{\textsf{het}}\)}\\ & \text{s.t.} && y\geq \frac{1}{D^i}\sum_{k\in K}D^{ik}\frac{z^{ik}}{\theta^i\cikdag(0)}-\frac{1}{D^{i'}}\sum_{k'\in K}D^{i'k'}\frac{z^{i'k'}}{\theta^{i'}c^{i'k'\dagger}(0)}, \notag\\ &{} && \quad\quad \quad  \forall i, i' \in I, \label{eq: HetTollsEquitable_y_1}\tag{\(\mathcal{P}^\ast_{\textsf{het}.a}\)}\\
   &{} && \sum_{i\in I}\sum_{k\in K}D^{ik} z^{ik} - \sum_{e\in E}p_ew_e^\dagger\geq T^\ast_{\textsf{het}}(\popuFlow^{\dagger{}}), \label{eq: HetTollsEquitable_opt}\tag{\(\mathcal{P}^\ast_{\textsf{het}.b}\)}\\
   &{} &&z^{ik} - \sum_{e\in r}(p_e^i+\gasPrice_e) \leq {\theta^i}\ell_r(w^{\dagger{}}),  \forall k\in K, r\in \paths^k, i\in I, \label{eq: HetTollsEquitable_eq}\tag{\(\mathcal{P}^\ast_{\textsf{het}.c}\)}\\
    &{} && p_e^i \geq 0, \forall e\in \edges, i \in I, \label{eq: HetTollsEquitable_nonneg}\tag{\(\mathcal{P}^\ast_{\textsf{het}.d}\)}
\end{align}
\end{subequations}}
where $T^\ast_{\textsf{het}}(\popuFlow^{\dagger})$ is the optimal value of the objective function of \eqref{eq: Het} associated with $\popuFlow^{\dagger}$. 
\end{proposition}

Propositions \ref{prop: HomHetDualOnly} and \ref{prop: HetTollsEquitable} show that $\phetopt$ can be computed using a \emph{three-step approach}: first, we compute the type-specific flow vector $\popuFlow^{\dagger}$ that induces the edge flow $\wdag$ while also minimizing the travel time difference among all traveler populations using \eqref{eq: cost_diff_min}. Second, we compute $T^\ast_{\textsf{het}}(\popuFlow^{\dagger})$ using \eqref{eq: Het} given $\popuFlow^{\dagger}$. Third, we compute $\phetopt$ using \eqref{eq: HetTollsEquitable}.

Finally, we discuss how to extend our approaches of computing $\phomopt$ and $\phetopt$ to incorporate the support constraints of the toll price. Previous studies \citep{hoefer2008taxing, bonifaci2011efficiency} showed that the problem of computing toll prices that satisfy support set constraints and also minimize the total travel time is NP hard even without considering heterogeneous \votTag\ of travelers or equity objectives. Here, we provide heuristics for computing the toll prices with support constraints. We evaluate the performance of our heuristics in terms of total travel time, equity, and welfare on the San Francisco Bay Area network in Sec. \ref{sec:method}. 

\paragraph{Heuristics for computing $\phomscopt$} 
We propose a two-step heuristic to compute \(\homosc\) by appropriately modifying the two-step method to compute \homo. 

We first solve the following linear program that adds the support constraints to \eqref{eq: HomToll}: 
  \begin{equation}\label{eq: HomscToll}\tag{\(\mathcal{P}_{\textsf{\homosc}}\)}
        \begin{aligned}
  T^\ast_{\homosc} = \max_{\tolls,z} &~ \sum_{i\in I}\sum_{k\in K}D^{ik} z^{ik} - \sum_{e\in \edges}\tolls_ew_e^{\dagger{}},\\ \text{s.t.} \quad &~  z^{ik} - \sum_{e\in r}(\tolls_e+\gasPrice_e) \leq \theta^i\ell_r(w^{\dagger{}}), \notag \quad \\&\quad \quad  \forall k\in K, r\in \paths^k, i\in I, \\
 {} &~ \tolls_e \geq 0, \quad \forall e\in E_T, \quad \tolls_e = 0, \quad \forall e\in E \setminus E_T.
\end{aligned}
\end{equation}

We note that the equilibrium edge load associated with any optimal solution of \eqref{eq: HomscToll}, say \(\hat{w}\), may not be equal to the socially optimal edge load $\wdag$. This is because the constraints that edges in $E \setminus E_T$ having zero tolls remove the dual constraints in \eqref{eq: HomDual_a} for edges in $E_T$. As a result, the primal and dual argument in the proof of Proposition \ref{prop: HomHetDualOnly} no longer holds, and thus the induced edge flow $\hat{w}$ may not be equal to $\wdag$.

Note that the optimal solution to \eqref{eq: HomscToll} will be non-unique. Therefore, inspired by \eqref{eq: HomTollsEquitable_obj}, we consider the following heuristic to incorporate both equity and welfare metric while also accounting for support constraints. 
Note that simply adding the support constraints in \eqref{eq: HomTollsEquitable_obj} could render the optimization problem infeasible as the optimal set of homogeneous tolls \(P^\ast_{\homo}\) need not have a solution that satisfies the support constraints. Particularly the constraint \eqref{eq: HomTollsEquitable_opt} would get violated. This is because \(T^\ast_{\homo} \geq T^\ast_{\homosc}\) as the constraint set of \eqref{eq: HomscTollsEquitable_obj} is contained in that of \eqref{eq: HomTollsEquitable_obj}. Therefore, we compute $\phomscopt$ as the optimal solution of the following linear program which adds support constraints to \eqref{eq: HomTollsEquitable_obj} while relaxing the constraint \eqref{eq: HomTollsEquitable_opt} by using \(T^\ast_{\homosc}\) instead of \(T^\ast_{\homo}\): 

{ \small\begin{subequations}\label{eq: HomscTollsEquitable}
\begin{align}
    &\min_{p,z, y} && y  + \frac{\lambda}{D} \sum_{i\in I}\sum_{k\in K}D^{ik}\frac{z^{ik}}{\theta^ic^{ik\dagger}(0)} \label{eq: HomscTollsEquitable_obj}\tag{\(\mathcal{P}^\ast_{\homosc}\)}\\ & \text{s.t.} && y\geq \frac{1}{D^i}\sum_{k\in K}D^{ik}\frac{z^{ik}}{\theta^i\cikdag(0)}-\frac{1}{D^{i'}}\sum_{k'\in K}D^{i'k'}\frac{z^{i'k'}}{\theta^{i'}c^{i'k'\dagger}(0)}, \notag\\ &{} && \quad\quad \quad  \forall i, i' \in I, \label{eq: HomscTollsEquitable_y_1}\tag{\(\mathcal{P}^\ast_{\homosc.a}\)}\\
   &{} && \sum_{i\in I}\sum_{k\in K}D^{ik} z^{ik} - \sum_{e\in E}p_ew_e^{\dagger{}}\geq T^\ast_{\homosc},{} \label{eq: HomscTollsEquitable_opt}\tag{\(\mathcal{P}^\ast_{\homosc.b}\)}\\
    &{} &&z^{ik} - \sum_{e\in r}(p_e+\gasPrice_e) \leq {\theta^i}\ell_r(w^{\dagger{}}), \forall k\in K, r\in \paths^k, i\in I, \label{eq: HomscTollsEquitable_eq}\tag{\(\mathcal{P}^\ast_{\homosc.c}\)}\\
 &{} && \tolls_e \geq 0, \quad \forall e\in E_T,  \tolls_e = 0, \quad \forall e\in E \setminus E_T. \label{eq: HomscTollsEquitable_nonneg}\tag{\(\mathcal{P}^\ast_{\homosc.d}\)}
\end{align}
\end{subequations}}

\paragraph{Heuristics for computing $\phetscopt$} 
The computation of $\phetscopt$ follows a three-step procedure, similar to that of $p^\ast_{\het}$ but restricting the set of allowable tolls to be zero on non-tollable edges, as done in \(\homosc\). 
First, we compute the population-specific flow vector $\popuFlow^{\dagger{}}$ that induces the congestion minimizing edge flow \(w^{\dagger{}}\) while also minimizes the average difference of travel time among all traveler populations using \eqref{eq: cost_diff_min}. Next, we add the support constraints to \eqref{eq: Het} to compute the optimal value \(T^\ast_{\hetsc}(f^{\dagger{}})\) as follows:
\begin{equation}\label{eq: HetscToll}\tag{\(\mathcal{P}_{\hetsc}\)}
    \begin{aligned}
    T^\ast_{\hetsc}(\popuFlow^{\dagger}) = \max_{p,z} & ~ \sum_{i\in I}\sum_{k\in K}D^{ik} z^{ik} - \sum_{e\in E}\sum_{i\in I}p_e^i\popuFlow_e^{\dagger i {}},\\ 
   \text{s.t.} & ~ z^{ik} - \sum_{e\in r}(p_e^i + \gasPrice_e )\leq \theta^i\ell_r(w^{\dagger{}}), \\ &\quad \quad \quad \forall k\in K, r\in \paths^k, i\in I, \\
  {} &~ \tolls_e^i \geq 0, \quad \forall e\in E_T \ \forall i\in I, \\&~ \tolls_e = 0, \quad \forall e\in E \setminus E_T \ \forall i\in I.
    \end{aligned}
\end{equation}
Analogous to the case $\homosc$, the equilibrium edge load associated with the optimal solution of \eqref{eq: HetscToll}, \(\hat{w}\), may not be equal to the socially optimal edge load $\wdag$ due to the added support constraints. 
We compute $\phetscopt$ as the optimal solution of the following linear program which adds support constraints to \eqref{eq: HetTollsEquitable} while relaxing the constraint \eqref{eq: HetTollsEquitable_opt} by using \(T^\ast_{\hetsc}\) instead of \(T^\ast_{\het}\): 
{\small
 \begin{subequations}\label{eq: HetscTollsEquitable}
\begin{align}
    &\min_{p,z, y} && y  + \frac{\lambda}{D} \sum_{i\in I}\sum_{k\in K}D^{ik}z^{ik}, \label{eq: HetscTollsEquitable_obj}\tag{\(\mathcal{P}^\ast_{\hetsc}\)}\\& \text{s.t.} && y\geq \frac{1}{D^i}\sum_{k\in K}D^{ik}\frac{z^{ik}}{\theta^i\cikdag(0)}-\frac{1}{D^{i'}}\sum_{k'\in K}D^{i'k'}\frac{z^{i'k'}}{\theta^{i'}c^{i'k'\dagger}(0)}, \notag\\ &{} && \quad\quad \quad  \forall i, i' \in I,\label{eq: HetscTollsEquitable_y_1}\tag{\(\mathcal{P}^\ast_{\hetsc.a}\)}\\
   &{} && \sum_{i\in I}\sum_{k\in K}D^{ik} z^{ik} - \sum_{e\in E}p_e^iw_e^{\dagger{}}\geq T^\ast_{\hetsc}, {} \label{eq: HetscTollsEquitable_opt}\tag{\(\mathcal{P}^\ast_{\hetsc.b}\)}\\
    &{} &&z^{ik} - \sum_{e\in r}(p_e^i+\gasPrice_e) \leq {\theta^i}\ell_r(w^{\dagger{}}),\notag \\ 
    &{} &&\quad \quad \forall k\in K, r\in \paths^k, i\in I, \label{eq: HetscTollsEquitable_eq}\tag{\(\mathcal{P}^\ast_{\hetsc.c}\)}\\
 &{} && \tolls_e^i \geq 0,~ \forall e\in E_T, i\in I, ~~  \tolls_e^i = 0,~ \forall e\in E \setminus E_T, i\in I.{} \label{eq: HetscTollsEquitable_nonneg}\tag{\(\mathcal{P}^\ast_{\hetsc.d}\)}
\end{align}
\end{subequations}
}

\section{Model calibration for the San Francisco Bay Area freeway network}\label{sec: ModelCalibrationBayArea}
In this section, we calibrate the non-atomic routing game model for the San Francisco Bay Area freeway network using the Caltrans Performance Measurement System (PeMS) dataset \footnote{available at \url{https://pems.dot.ca.gov/}}, American Community Survey (ACS) dataset \footnote{available at \url{https://www.census.gov/programs-surveys/acs}} and Safegraph neighborhood patterns dataset from 2019 \footnote{This dataset was available for public use at \url{https://www.safegraph.com} till 2021 and is now commercially available}. In Sec. \ref{ssec: IntroDataset}, we briefly describe each dataset. We subsequently present the calibration of the Bay Area transportation network, the demand of each population type in Sec. \ref{ssec: flow_estimation}, and the \votTag\ parameters in Sec. \ref{vot_estimation}.

\subsection{Datasets}\label{ssec: IntroDataset}

\paragraph{Caltrans PeMS Dataset} The Caltrans PeMS dataset is based on measurements taken from loop detectors placed on a network of freeways and bridges in California. Our dataset is taken from district 4, which covers the entire San Francisco Bay Area. This dataset provides hourly flow counts and average vehicle speeds measured by each loop detector placed along the freeways. We use this dataset to calibrate the latency functions of edges (See Sec. \ref{ssec: flow_estimation} for detailed discussion).

\paragraph{American Community Survey (ACS) Dataset} 
The ACS dataset is collected by the US Census Bureau to record demographic and socioeconomic information. We use the information from Means of Transportation (2019) entry in ACS, which provides information of commuters' mode choices (percentage of driving population), employment, and household income. The dataset is collected at the zip-code level for the entire United States. 

\paragraph{Safegraph Neighborhood Patterns Dataset} This dataset records the aggregate mobility pattern using the data collected from ~40 million mobile devices in the US. This dataset estimates the commuting pattern by counting the number of mobile devices that travel from one census block group (CBG) to another CBG and dwell for at least 6 hours between 7:30 am and 5:30 pm Monday through Friday. We use this dataset in conjunction with ACS and the Safegraph datasets to estimate the demand of driving commuters between each o-d pair in the network within each income level (See Sec. \ref{ssec: flow_estimation} for detailed discussion).

\subsection{The San Francisco Bay Area freeway network}\label{ssec: flow_estimation}
\begin{figure}[h!]
    \centering
     \begin{minipage}{0.45\columnwidth} 
        \centering
\includegraphics[width=0.9\textwidth, height=7cm]{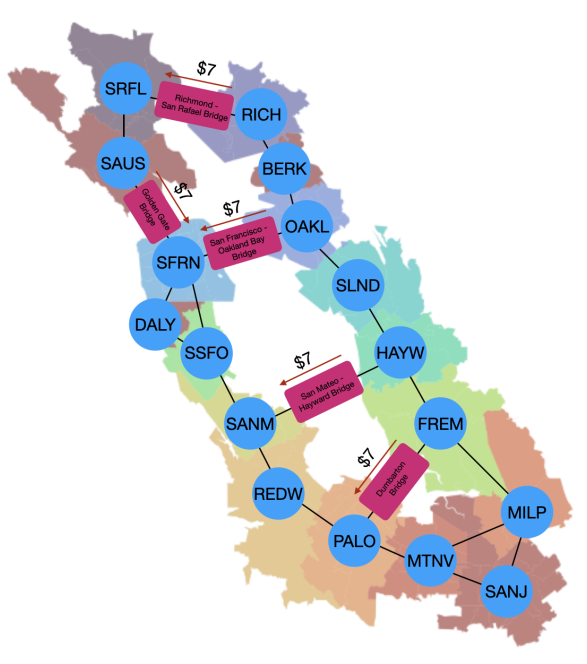} 
    \end{minipage}
     \hspace{0.05\columnwidth} 
    \begin{minipage}{0.45\columnwidth} 
        \centering
        \begin{tabular}{ l r r }
      \toprule
      Node & Abbr \\
      \midrule
       San Rafael & \SNRF \\ 
Richmond & \RICH \\
Oakland & \OAKL \\ 
San Francisco & \SANF \\
San Leandro & \SANL \\ 
Hayward & \HAYW \\ 
South SF & \SSFO \\ 
Fremont & \FREM \\ 
San Mateo & \SANM \\ 
Redwood City & \REDW \\ 
Palo Alto & \PALO \\ 
Milpitas & \MILP \\ 
Mountain View & \MOUN \\
San Jose & \SANJ \\ 
Sausalito & \SAUS \\ 
Daly City & \DALY \\ 
Berkeley & \BERK\\
      \bottomrule
    \end{tabular}
    \end{minipage}%
   \caption{Bay Area transportation network with tolled bridge segments. Different colors on the map represent the boundaries of cities. The table contains the names of the nodes in map along with abbreviations.}
       \label{fig:node_map}
\end{figure}

We represent the San Francisco Bay Area using a network with 17 nodes (see Fig. \ref{fig:node_map}). Each node represents a major city, and the edges are the major freeways connecting these cities. Among these edges, five of them are bridges: the Golden Gate Bridge, the Richmond-San Rafael Bridge, the San Francisco-Oakland Bay Bridge, the San Mateo-Hayward Bridge, and the Dumbarton Bridge. They are represented as the magenta boxes in Figure \ref{fig:node_map}. In 2019, a flat toll of \$7 is imposed for a single crossing on each bridge in the direction denoted in Figure \ref{fig:node_map}.

\paragraph{Demand estimate}
We categorize the driving population into three distinct segments based on their \votTag, namely \emph{low, middle, and high} \votTag. The determination of the fraction of driving population in each of these categories relies on the \emph{Means of Transportation} dataset from ACS. Specifically, we assign a traveler to the (a) low \votTag\ category if their annual individual income is less than $\$ 25,000$, to the (b) middle \votTag\ category if their annual individual income falls within the range of $\$ 25,000$ to $\$ 65,000$, and to the (c) high \votTag\ category if their annual individual income exceeds $\$ 65,000$.

\begin{figure*}[!t]
\centering
\subfloat[Distribution of traveler types based on their origin (home) nodes]{\includegraphics[width=0.45\linewidth]{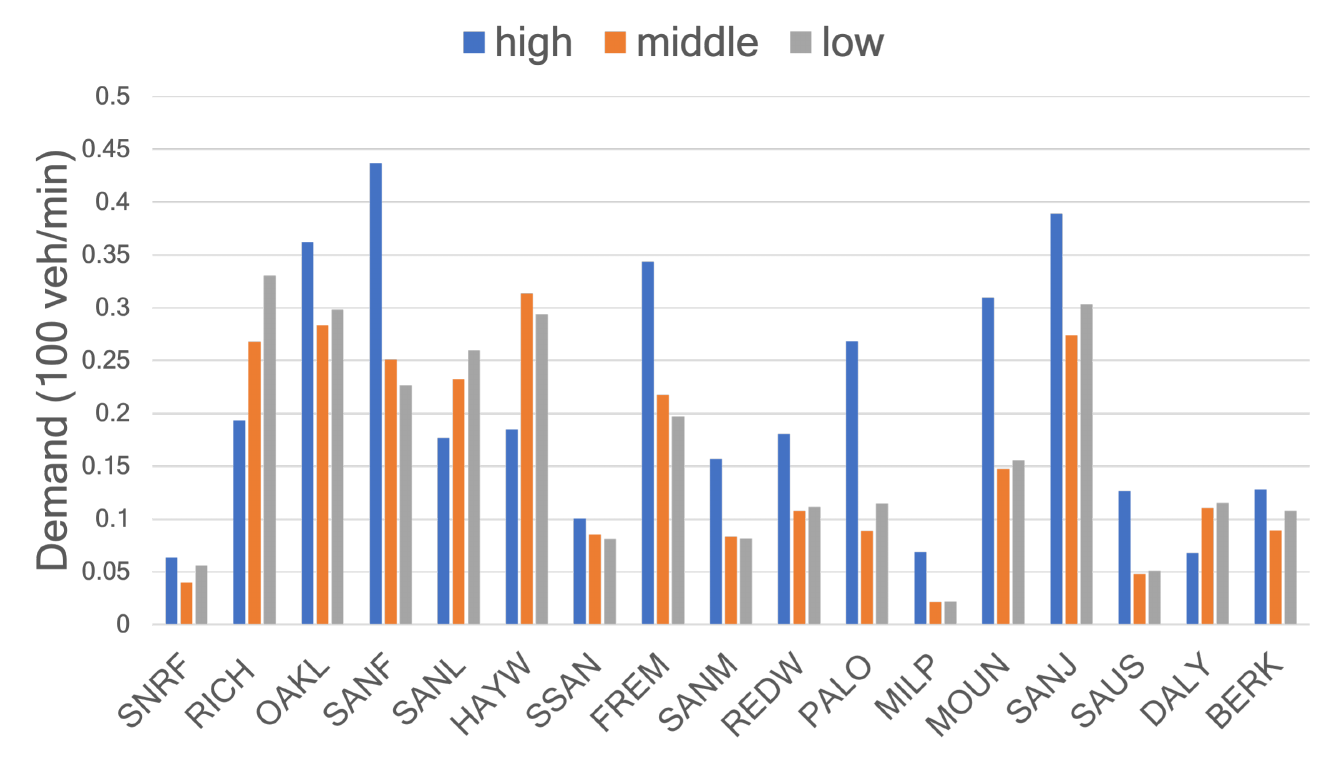}%
\label{fig: DemandOrigin}}
\hfill
\subfloat[Distribution of traveler types based on their destination (work) nodes.]{\includegraphics[width=0.45\linewidth]{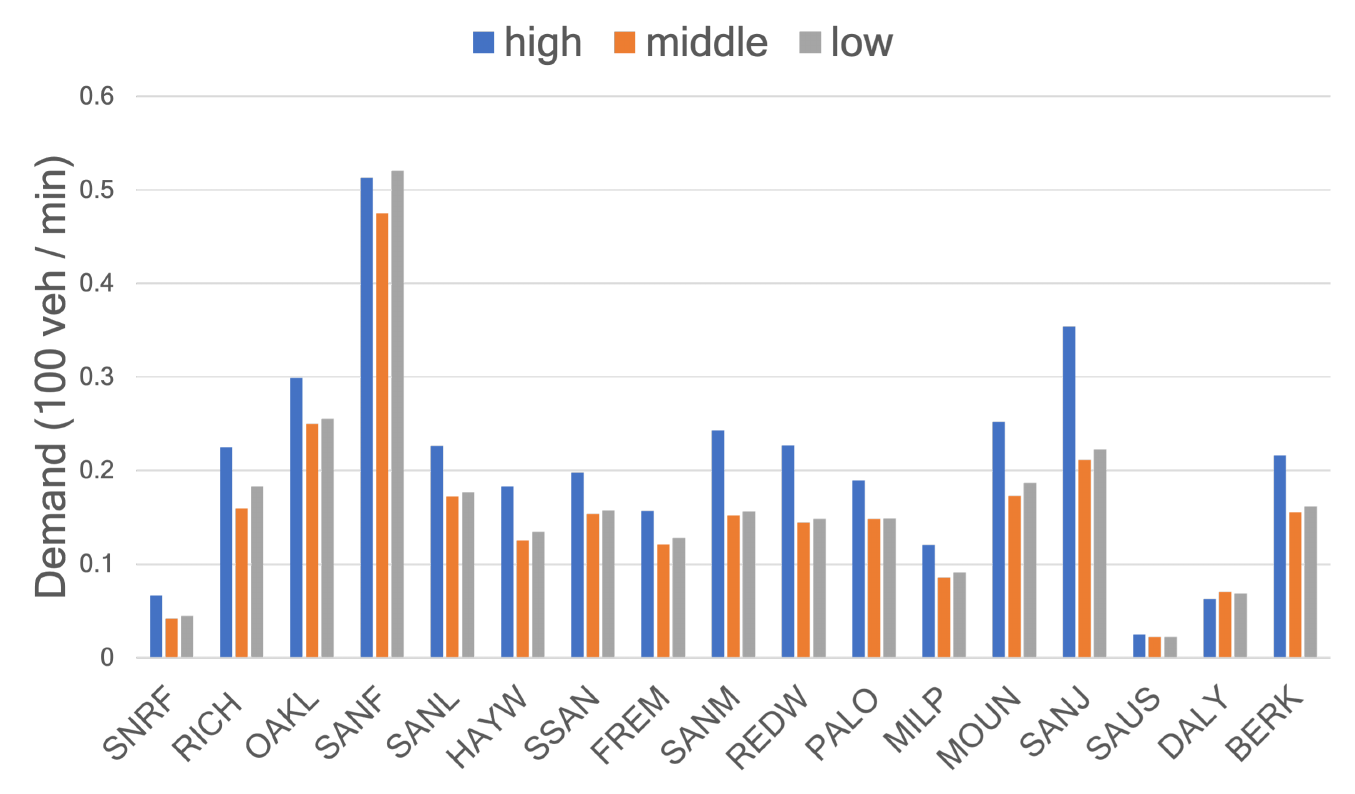}%
\label{fig: DemandDestination}}
\caption{Distribution of origin and destination traveler demands. }
\label{fig:three graphs}
\end{figure*}

\begin{figure*}[!t]
\centering
\subfloat[Home Location]{\includegraphics[width=0.24\linewidth]{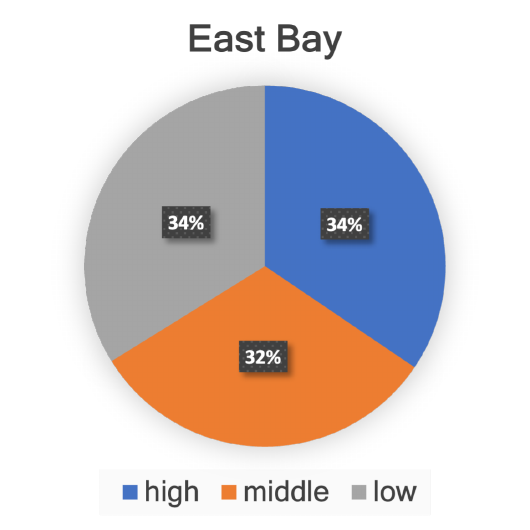}%
\label{fig:EastBayOrigin}}
\hfill
\subfloat[Home Location]{\includegraphics[width=0.24\linewidth]{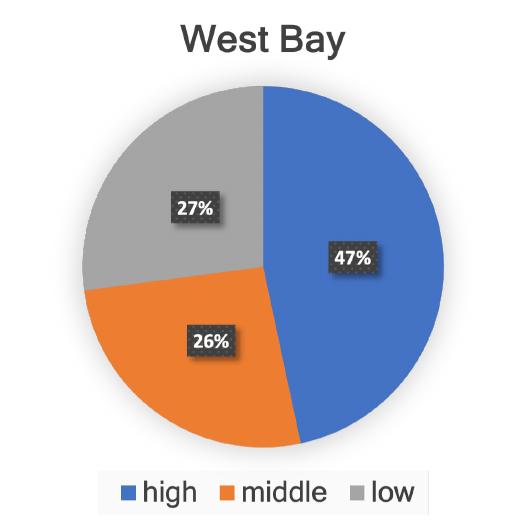}%
\label{fig:WestBayOrigin}}
\hfill
\subfloat[Work Location]{\includegraphics[width=0.24\linewidth]{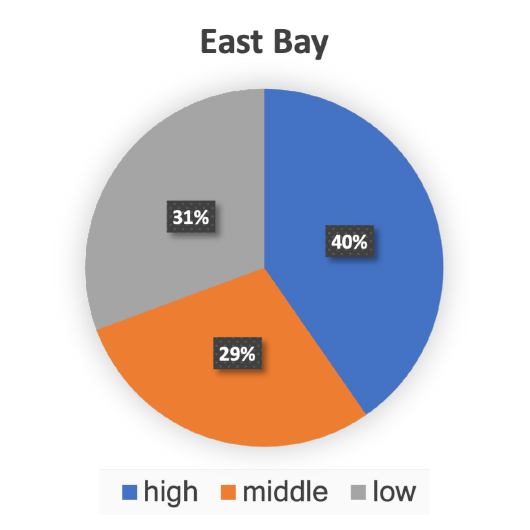}%
\label{fig:EastBayDestination}}
\hfill
\subfloat[Work Location]{\includegraphics[width=0.24\linewidth]{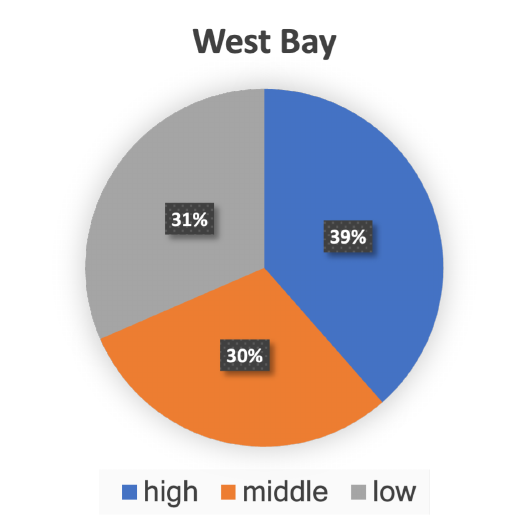}%
\label{fig: WestBayDestination}}
\caption{{Distribution of home and work demands across high, middle, and low income levels aggregated over the two sides of the bay area.}}
 \label{fig:DistributionDemandEastBayWestBay}
\end{figure*}

Figure \ref{fig:three graphs} provides a visual representation of the distribution of traveler demand to and from each node in the network, stratified by \votTag. Note that this demand specifically pertains to inter-node travel, with within-node demand excluded from the analysis. We find that approximately 40\% of travelers are high willingess-to-pay, and 30\% of travelers are of middle and low \votTag, each. 
In Figure \ref{fig: DemandOrigin} (resp. Figure \ref{fig: DemandDestination}), we present the distribution of traveler demand based on their home (resp. work) location.  Around \(55\%\) of traffic emerges from relatively few nodes on the East Bay such as \RICH, \OAKL, \SANL, \HAYW, \FREM, \SANJ. Moreover, around \(40\%\) of traffic  has a work destination in one of the four nodes \SANF, \PALO, \MOUN, and \OAKL.    
Notably, there exists substantial heterogeneity in both the home and work locations of different traveler types, as can be observed by comparing the distribution of demands in Figure \ref{fig:three graphs} to the distribution of median income found in Figure \ref{fig: MedianIncome}. For instance, nodes such as \RICH, \HAYW, \SANL, and \DALY~ are predominantly inhabited by a higher number of low \votTag~ travelers, while nodes such as \PALO, \OAKL, \SANF, \FREM, and \SAUS~ are predominantly inhabited by high \votTag\ travelers.  It is interesting to note that on most of the nodes the demographics of incoming traffic predominantly comprise high \votTag\ travelers.  Additionally,
as can be seen in Figure \ref{fig:DistributionDemandEastBayWestBay}, 
high-income travelers make up a large fraction demand that originates in the West Bay, as well as of the work location demand on both the East and West Bay.

Next, we describe the approach used to compute the daily demand of different types of travelers traveling between different o-d pairs during January 2019-June 2019. There are three main steps to our approach: first, we obtain an estimate of the relative demand of travelers traveling between different zip-codes in the Bay Area by using the Safegraph dataset. Particularly, for every month, the Neighborhood Patterns data in the Safegraph dataset provides the average daily count of mobile devices that travel between different census block groups (CBGs) during the work day, which is then aggregated to obtain the relative demand of travelers traveling between different zip codes. After accounting for the sampling bias induced due to the randomly sampled population across the United States, we calibrate demands by using the ACS dataset which provides the income-stratified driving population in every zip code. Finally, to obtain an estimate of daily variability in demand we further augment the demand data with the PeMS dataset by adjusting for daily variation in the total flow on the network in every month. The details of demand estimation are included in \cite{maheshwari2024congestion}.

\paragraph{Calibrating the edge latency functions} We calibrate the latency functions of each edge of the Bay Area freeway network shown in Figure \ref{fig:node_map}.  We adopt the Bureau of Public Roads (BPR) function proposed by the Federal Highway Administration (FHA) \citep{manual1964application},defined as \(
    \latency_e(\edgeFlow_e) = a_e+b_e w_e^4,\) for every \(e\in E,\) 
where \(a_e\) represents the free-flow travel time (i.e. latency with zero flow) of edge $e$ and \(b_e\) is the slope of congestion.

We compute the average driving time of each edge during the morning rush hour (6am to 12pm) on each workday from January 1, 2019 to June 30, 2019 using the speed and distance data from the PeMS dataset. We denote the set of all days as \(\mathcal{T}\), the travel time and traffic flow of each edge \(e\in E\) on day \(t\in \mathcal{T}\) as \(\hat{\ell}_e^{t}\) and \(\hat{w}_e^t\), respectively. 
The details of computing \(\left(\hat{\ell}_e^{t}, \hat{w}_e^t\right)_{t \in \mathcal{T}}\) are provided in \cite{maheshwari2024congestion}. We estimate the free-flow travel time $a_e$ of each $e\in E$ using the average travel time of edge $e$ computed from the PeMS dataset at 3am, when the traffic flow is approaching zero. We denote the estimated value of \(a_e\) as \(\hat{a}_e\) for each $e \in E$. We next estimate the slope $b_e$ of each edge $e \in E$ using an ordinary least squares regression. In particular, the estimate $\hat{b}_e$ is solved as the minimizer of the following convex program: for every \(e\in E,\) \(\hat{b}_e = \underset{b_e\in \mathbb{R}}{\arg\min} \sum_{t \in \mathcal{T}} \|\hat{\ell}_{e}^{t} - \hat{a}_e- b_e \cdot (\hat{w}_{e}^{t})^{4}\|^2.\)

\subsection{Estimating the \votTag\ parameters}\label{vot_estimation}
We formulate the problem of estimating the \votTag\ parameters as an inverse optimization problem. Specifically, the optimal estimate of \votTag\ parameters corresponding to the three types of travelers, \(\vot^{H*}, \vot^{M*}, \vot^{L*}\), are the ones that minimize the difference between the observed flows on each edge of the network and the corresponding equilibrium edge flows. That is,  
\begin{subequations}\label{eq: VOTCompute}
\begin{align}
    \vot^{*}_H, \vot^{*}_M, \vot^{*}_L &= \argmin_{\substack{\vot^H, \vot^M, \vot^L }} \quad  \sum_{\days \in \mathcal{T}}\sum_{\edge \in \edges}(\edgeFlowObs_{\edge}^{\days} - \edgeFlow_{\edge}(\pathFlow^{\days}))^{2} \notag\\
    \text{s.t.} \quad & \pathFlow^{\days} \in \underset{q \in \mathcal{Q}(D^t)}{\arg\min}\ \Phi(q,p, \vot)\ \quad  \forall \days \in \mathcal{T} \label{subeq:potential}, \\
    & \text{\(\edgeFlow_{\edge}(q^t)\) is given by \eqref{eq: EdgeFlows},}
     \\
    & \text{\(\mathcal{Q}(D^t)\) is given by \eqref{eq: SetFeasible},}
\end{align}
\end{subequations}
where \(p\) is the toll price vector in 2019 (i.e. $\$ 7$ on each bridge, and $\$ 0$ for the remaining edges), $\edgeFlowObs_{\edge}^{\days}$ is the observed edge flow on each edge $e \in E$ and each day $t \in \mathcal{T}$ computed using the PeMS dataset, and $D^t$ is the estimated demand vector of each day $t$ computed using the ACS and Safegraph datasets.

Directly solving \eqref{eq: VOTCompute} is challenging due to the non-linearity of the edge latency function and the potential function in \eqref{subeq:potential}. We compute the estimates using grid search: we construct a grid of \votTag, where the granularity of each of $\theta^H, \theta^M, \theta^L$ is $\$ 5$ per hour. We also assume that the maximum value of \votTag\ is $\$ 100$ per hour and the minimum is $\$ 0$ per hour. Therefore, we define the set of all possible parameter values as \(\Theta := \{0, 5, 10, 15, \dots, 100\}^{3}\).
For each \(\vot = (\vot^{H}, \vot^{M}, \vot^{L}) \in \Theta \), we compute the equilibrium flow \(\pathFlow_{\days}\) for every \(t\in  \mathcal{T}\) and compute the total squared error as in the objective function of \eqref{eq: VOTCompute}. The optimal parameter \(\theta^\ast\) is the one that minimizes the total squared error. We obtain: 
\begin{equation*}
    \vot^{*} = (\theta^{L*}, \theta^{M*}, \theta^{H*}) = (\$10/\text{hour},\ \$30 /\text{hour},\ \$70 /\text{hour}).
\end{equation*}

Our estimate $\theta^*$ is consistent with the observations reported in prior works, which show that the \votTag\ values typically lie between \(60\%-100\%\) of the average hourly income of the population (\cite{palmquist2007measuring,athira2016estimation,meunier2015value}). 

\begin{figure}
    \centering
\includegraphics[width=0.35\textwidth]{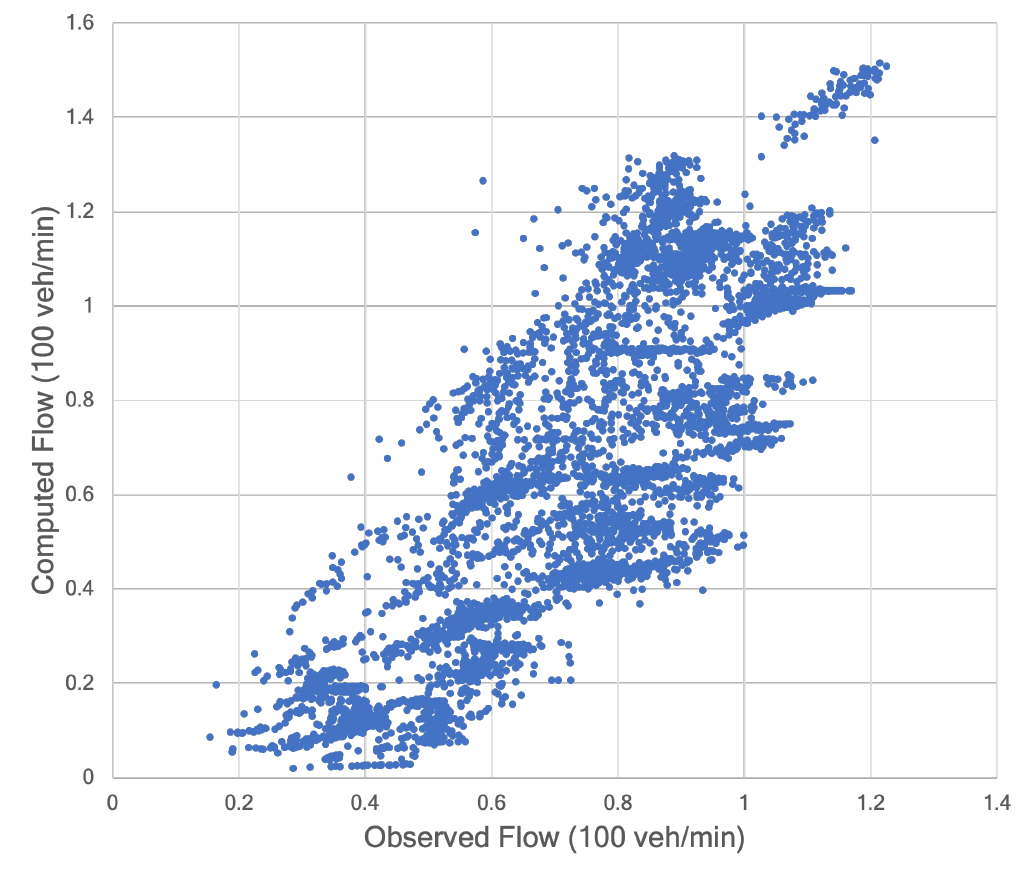}
\caption{Observed and computed equilibrium edge flow. 
} 
    \label{eq_vs_observed_flow}
\end{figure}

Furthermore, as a robustness check, we plot the equilibrium edge flow \(\edgeFlow_{\edge}(q^{t\ast})\) and observed edge flow \(\edgeFlowObs_{\edge}^{\days}\) for every \(e\in E, t\in \mathcal{T}\) in Figure \ref{eq_vs_observed_flow}. Each dot in this figure represents the flow on an edge $e \in E$ on a single day $t \in \mathcal{T}$.
Overall, the dots are distributed along the diagonal of the plot indicating that the our computed equilibrium edge flow are relatively consistent with the observed edge flow subject to noise in time costs and demand fluctuations.

\section{Efficiency and equity analysis of congestion pricing schemes}\label{sec: Analysis}
Our goal in this section is three fold. First,
we analyze the congestion levels induced at equilibrium due to current congestion pricing scheme, \curr, and identify corridors in the Bay Area which are congested. Next, using the computational method introduced in Section \ref{sec:method} and the calibrated model of San Francisco Bay area freeway network in Section \ref{sec: ModelCalibrationBayArea},  we compute the toll values under the congestion pricing schemes \homo, \het, \homosc, and \hetsc. 
Finally, we compare different congestion pricing schemes in terms of efficiency and equity of travel cost, and also in terms of overall revenue generated at equilibrium. 

\subsection{Congestion under the current congestion pricing scheme (\curr)}
Here, we analyze the congestion levels induced at equilibrium under the current congestion pricing scheme, \curr, which imposes a uniform toll of \$7 on each of the five bridges in the Bay Area, namely on the Richmond-San Rafael Bridge (\RICH-\SNRF), San Francisco-Oakland Bay Bridge (\OAKL-\SANF), Golden Gate Bridge (\SAUS-\SANF), San Mateo-Hayward Bridge (\HAYW-\SANM), and Dumbarton Bridge (\FREM-\PALO). 

\begin{figure*}[!t]
\centering
\subfloat[Proportional travel time increase under $\curr$  (normalized by free flow travel time).]{\includegraphics[width=0.45\linewidth]{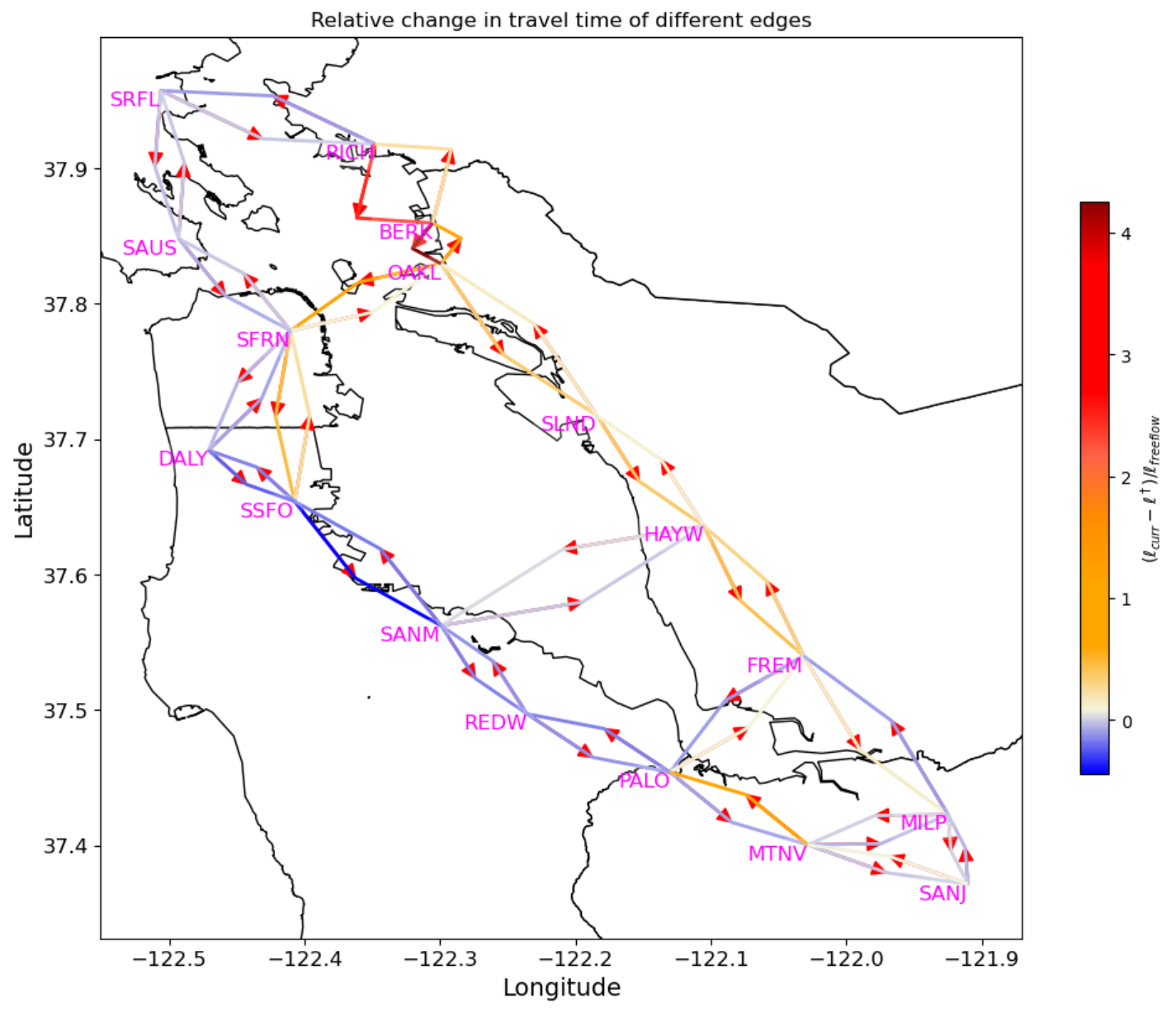}
\label{fig:relativeLatency}}
\hfill
\subfloat[Difference between equilibrium flow induced by \curr\ and optimal flow.]{\includegraphics[width=0.45\linewidth]{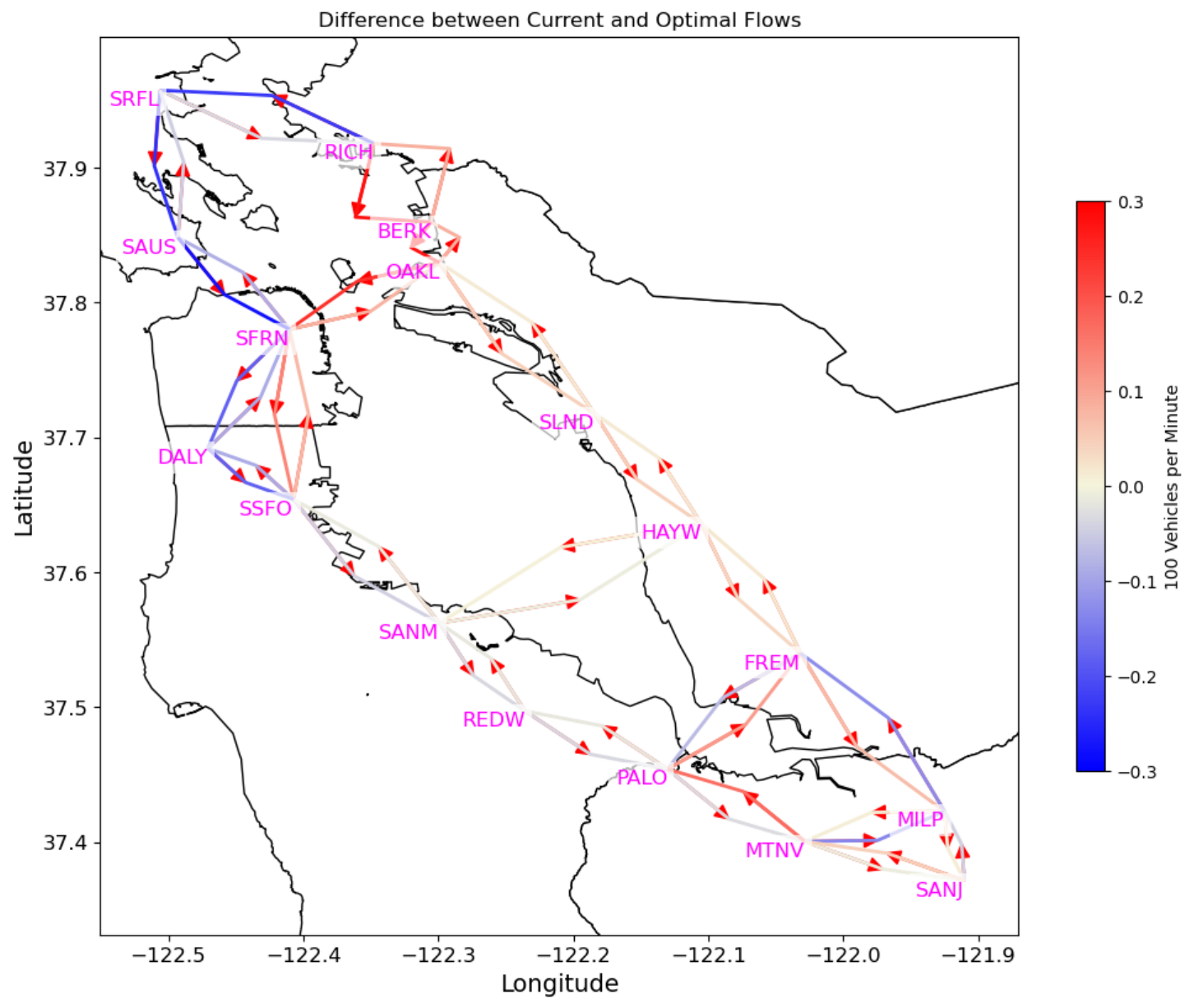}%
\label{fig:currDiffopt}}
\caption{Current congestion pricing scheme}
\label{fig:CongestionPatternCurrent}
\end{figure*}

Figure \ref{fig:relativeLatency} depicts the difference between the equilibrium travel time given \curr\ and the congestion minimizing travel time (normalized by free flow travel time on every edge). We observe that edges on the eastern corridor (connecting nodes \RICH-\BERK-\OAKL-\SANL-\HAYW-\FREM) are over-congested. Meanwhile, the edges on the western corridor (connecting nodes \SRFL-\SAUS-\SFRN-\DALY-\SSFO-\SANM-\REDW) are relatively less congested. Furthermore, we observe that amongst all bridges the Bay Bridge (\OAKL-\SANF) is also most congested, which is   
consistent with several prior studies \citep{nakamura2002congestion, barnes2012impact, gonzales2015empirical}. 
Additionally, Figure \ref{fig:currDiffopt} presents the difference in the edge flows induced at equilibrium with that of socially optimal edge flows.  We observe that in order to reduce the overall congestion we need to ensure that 

\begin{enumerate}[left= 1cm, leftmargin=*]
    \item[(R1)]\label{item: R1} 
    the travelers using the edges in the corridor \RICH-\BERK-\OAKL-\SANF\ (resp. \SANF-\OAKL-\BERK-\RICH)
    are incentivized to 
 use the edges in the corridor \RICH-\SNRF-\SAUS-\SANF\ (resp. \SANF-\SAUS-\SNRF-\RICH). 
 \item[(R2)]\label{item: R2} the travelers using the edges in the corridor \SANF-\SSAN\ are incentivized to use the corridor \SANF-\DALY-\SSAN. 
    \item[(R3)]\label{item: R3} the travelers using the eastern corridor \MILP-\FREM-\HAYW-\SLND-\OAKL\ are diverted to use the western corridor \MOUN-\PALO-\REDW-\SANM-\SSFO\ by suitably incentivizing them to use the Dumbarton Bridge or the San Mateo-Hayward Bridge.
\end{enumerate}

Furthermore, we note that the average travel cost\footnote{It can be shown that the average travel cost experienced by travelers is independent of the route flows on the network and is only dependent on the equilibrium edge flows, which are unique as shown in Proposition \ref{prop: PotentialFunction}.} (the sum of the travel time cost and the equivalent time cost of the monetary expense as in \eqref{eq: TotalCost}) experienced by different types of travelers at equilibrium is unequal in \curr. Specifically, low \votTag\ travelers bear the travel cost of approximately 
91 minutes, while high and middle \votTag\ travelers face costs of 
61 and 
68 minutes, respectively. Moreover, as indicated in Table \ref{tab:equity_Curr_Pricing}, this unequal distribution of travel time persists not only on average but also when examined across different threshold levels of travel cost.

\begin{table}
    \centering
    \begin{tabular}{c|c|c|c}
      Travel Cost    & Low (\%) &   Middle (\% ) & High (\%) \\ 
      \hline 
      \hline 
    \(\geq\) 60 minutes & 69 & 55 & 46 \\ 
    \(\geq\)  90 minutes & 51 & 31 & 28\\ 
   \(\geq\)   120 minutes & 32 & 13 & 12\\ 
\(\geq\)   150 minutes & 17 & 1 & 1\\
\hline 
    \end{tabular}
    \caption{Fraction of low, middle and high \votTag\ travelers that incur total cost (in minutes) more than stated threshold at equilibrium.}
    \label{tab:equity_Curr_Pricing}
\end{table}

To summarize, we observe that the current congestion pricing scheme implemented the Bay area does not result in efficient allocation of traffic on the network. Additionally, it also leads to unequal distribution of travel cost across different types of travelers. 

\subsection{Toll values under different congestion pricing schemes}
Here, using the calibrated model of the Bay area obtained in Section \ref{sec: ModelCalibrationBayArea}, we present the computed values of tolls on various edges of the Bay area network under different congestion pricing schemes (namely, \homo, \het, \homosc, \hetsc) obtained using the computational methodology presented in Section \ref{sec:method}.

\begin{figure*}[!t]
\centering
\subfloat[Tolls under \homo]{\includegraphics[width=0.45\linewidth]{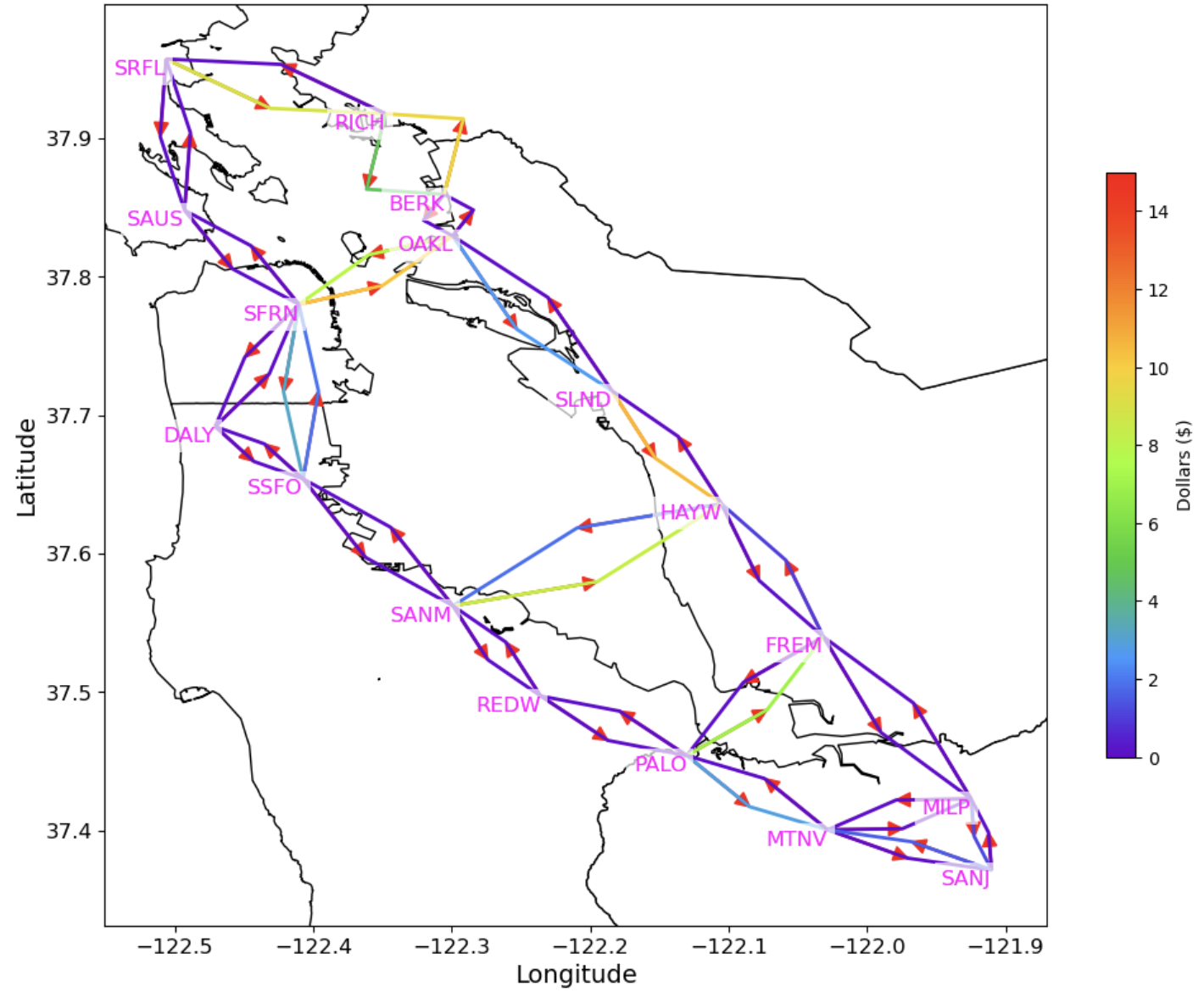}
\label{fig:tolls_hom}}
\subfloat[Tolls under \het\ for low type]{\includegraphics[width=0.45\linewidth]{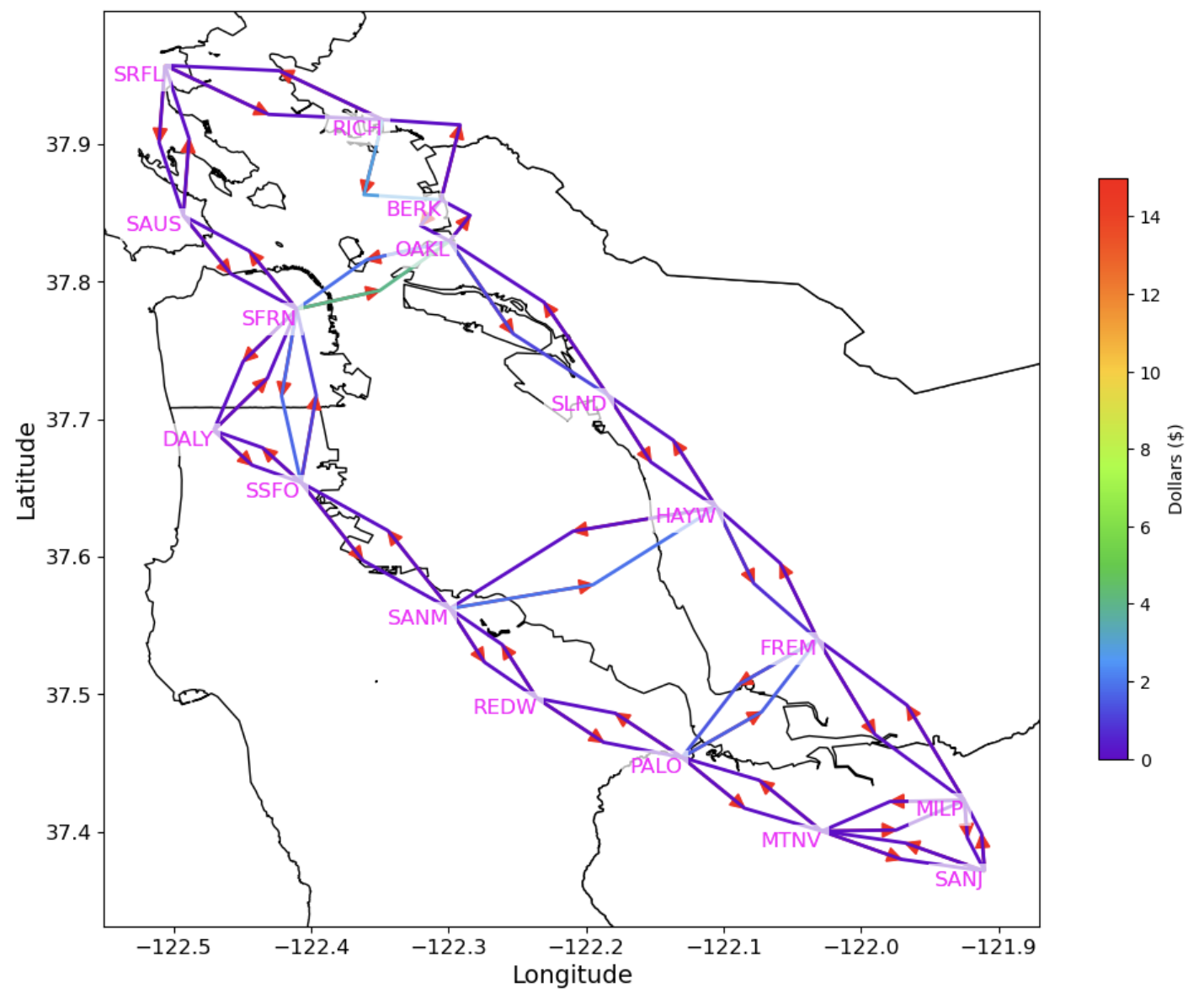}%
        \label{fig:tolls_het_low}
}

\subfloat[Tolls under \het\ for middle type]{\includegraphics[width=0.45\linewidth]{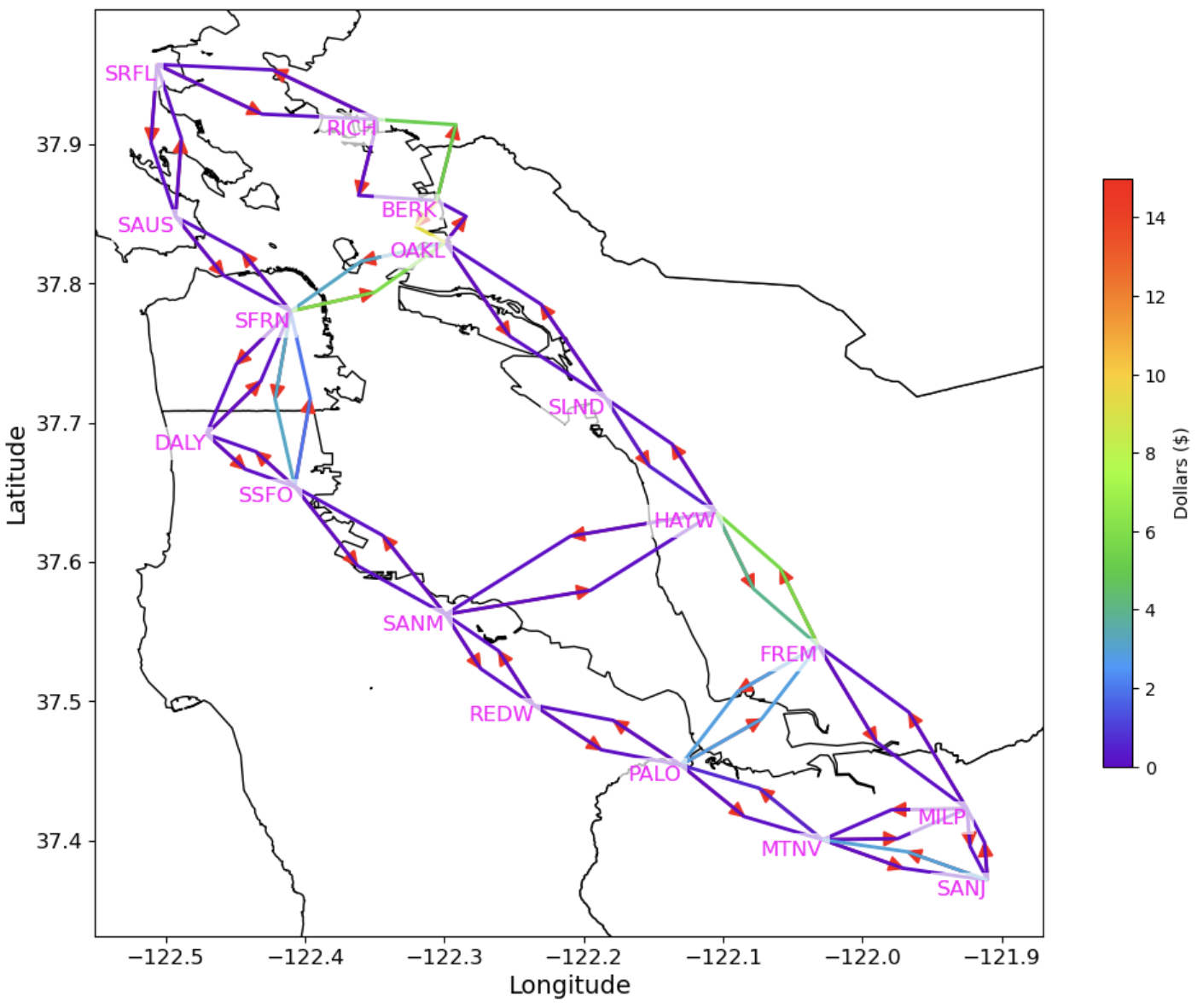}
\label{fig:tolls_het_middle}}
\subfloat[Tolls under \het\ for high type]{\includegraphics[width=0.45\linewidth]{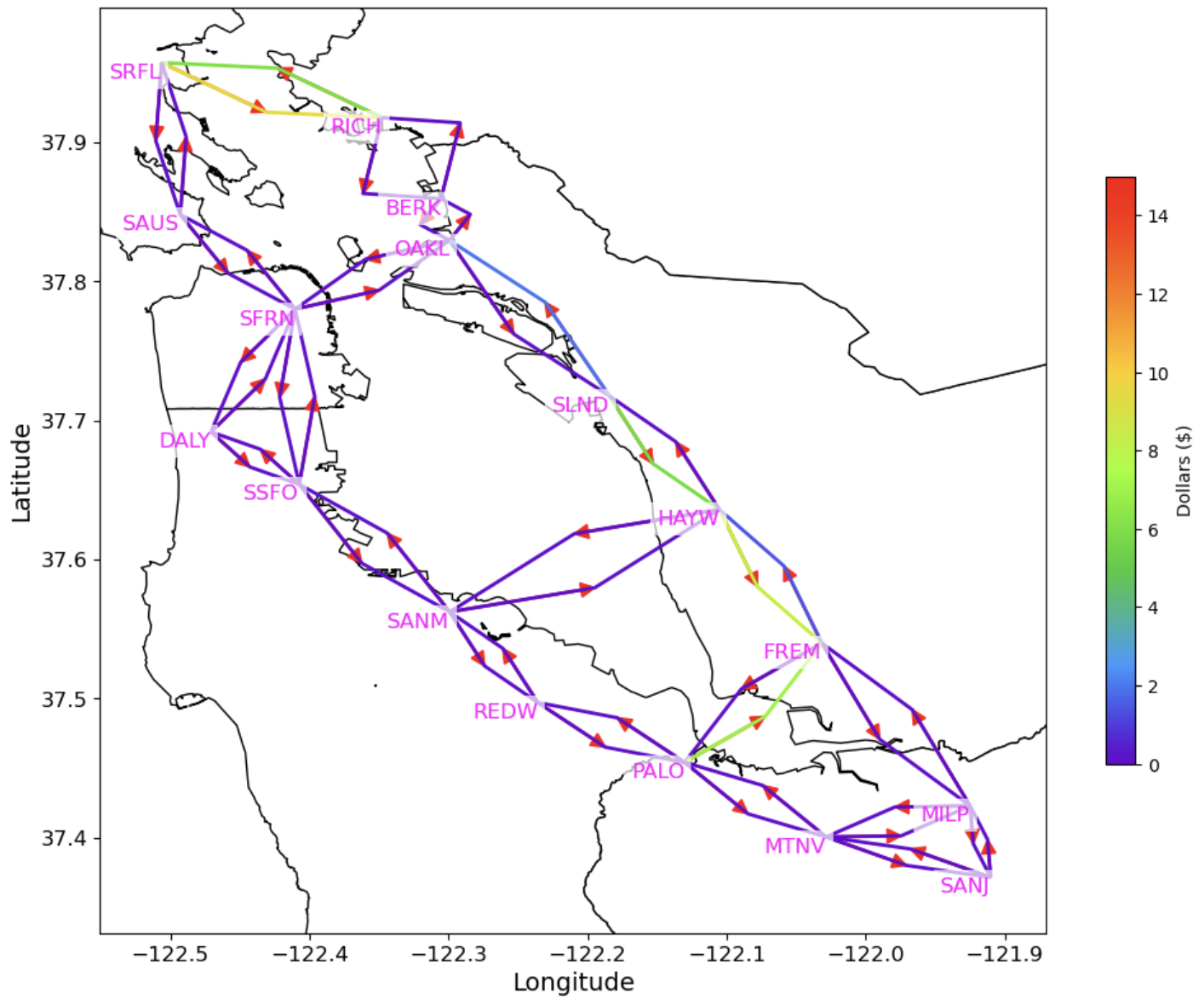}%
        \label{fig:tolls_het_high}
}

\subfloat[Tolls under \homosc]{\includegraphics[width=0.45\linewidth]{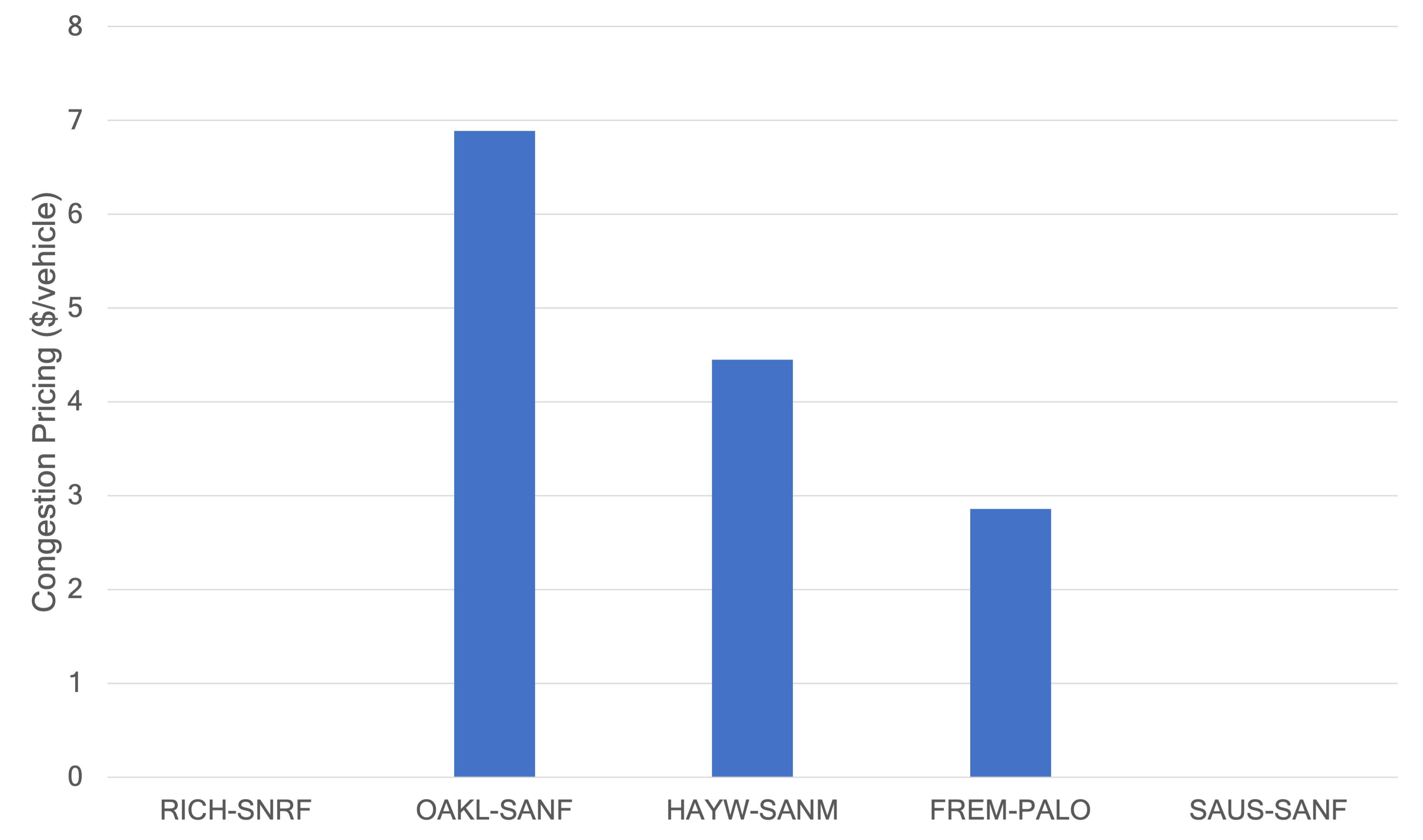}
\label{fig:tolls_homsc}}
\subfloat[Tolls under \hetsc]{\includegraphics[width=0.45\linewidth]{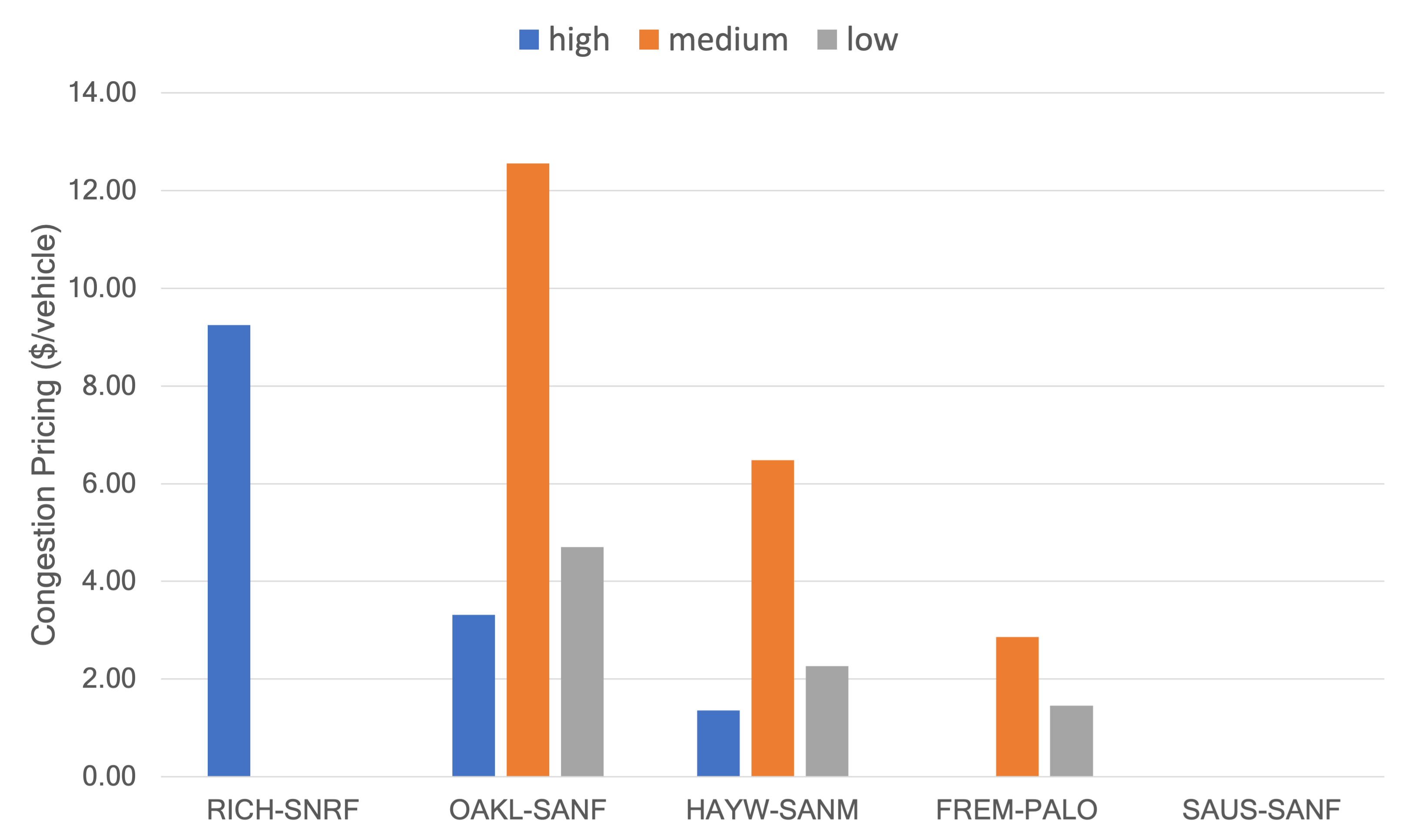}%
        \label{fig:tolls_hetsc_all}
}
\caption{Toll values under congestion pricing schemes \homo, \het, \homosc, and \hetsc.}
\label{fig:toll_vals_all_congestion_pricing}
\end{figure*}

Figure \ref{fig:tolls_hom} presents the toll values computed under \homo\ by solving \eqref{eq: HomTollsEquitable_obj}. Figures \ref{fig:tolls_het_low}-\ref{fig:tolls_het_high} present the toll values for low, middle, and high \votTag\ travelers under \het\ by solving \eqref{eq: HetTollsEquitable}. Figure \ref{fig:tolls_homsc} presents the toll values computed under \homosc\ by solving \eqref{eq: HomscTollsEquitable_obj}. Figure \ref{fig:tolls_hetsc_all} further presents the toll values for low, middle, and high \votTag\ travelers under \hetsc\ by solving \eqref{eq: HetscTollsEquitable_obj}. 
To compute all of these toll values, we choose \(\lambda = 20\) in \eqref{eq: HomTollsEquitable_obj}, \eqref{eq: HetTollsEquitable}, \eqref{eq: HomscTollsEquitable_obj}, and \eqref{eq: HetscTollsEquitable_obj}. 
This choice of parameter \(\lambda\) ensures that the numerical value of the average welfare metric and the equity metric in these optimization problems are of the same order of magnitude.

Note that in \homo\ and \het, on all the bridges, tolls in the east-to-west direction are lower than tolls in the west-to-east direction. This is in contrast to \curr, where the west-to-east direction is not tolled at all on any bridge and only the east-to-west direction is tolled at a flat rate of $\$7$ (refer Figure \ref{fig:node_map}). Given that the western corridor is less congested than the eastern corridor in \curr\ (refer Figure  \ref{fig:relativeLatency}), such tolling is useful to efficiently redistribute traffic in the network.  Furthermore, note that in all of the congestion pricing schemes we compute, unlike \curr, the Golden Gate Bridge (\SAUS-\SANF) is not tolled at all. This choice ensures that more travelers in the eastern corridor, particularly in nodes such as \RICH~ and \BERK~ are able to reach nodes in the west, particularly \SANF, through Golden-Gate bridge instead of Bay-bridge (\OAKL-\SANF). 

\subsection{Discussion on efficiency, equity and revenue generation}
In this subsection, we compare the effectiveness of \curr, \homo, \homosc, \het\ and \hetsc\ in  terms of efficiency (the average travel time per traveler), equity (average increase in travel cost in comparison to no tolls), and revenue generation (the total toll revenue generated by these schemes). Additionally, we also compare these pricing schemes with the scenario when no toll is implemented (denoted \zero). 

\subsubsection{Efficiency and Equity Considerations.}\label{ssec: EfficiencyConsiderations}
\begin{figure}[htbp]
    \centering
    \includegraphics[width = 0.6\textwidth]{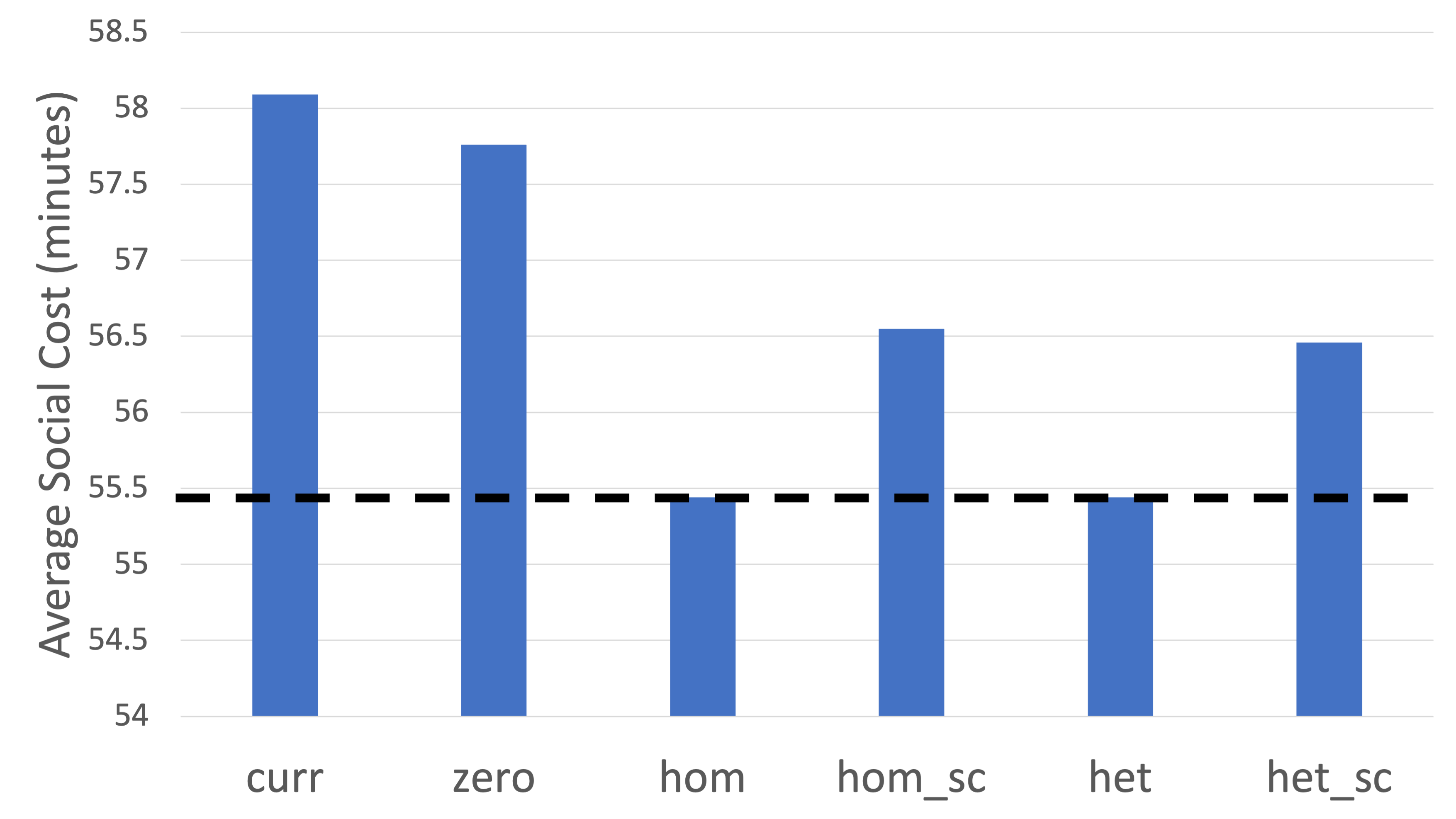}
    \caption{Comparison of average social cost per traveler for \curr, \zero, \homo, \homosc, \het, and \hetsc. Here, the dashed line represent congestion minimizing cost computed by solving \eqref{eq: SocOptProblem}.} 
    \label{fig: SocCost}
\end{figure}

Figure \ref{fig: SocCost} represents the average travel time experienced by travelers under different congestion pricing schemes.  

As expected from Proposition \ref{prop: HomHetDualOnly}, the congestion pricing schemes \homo\ and \het\ achieves the minimum congestion levels on the network.  Additionally, we note that \homosc\ and \hetsc\ do not achieve the minimum congestion level due to the support constraints. 
Furthermore, it's noteworthy that \hetsc\ results in a slightly improved average travel time compared to \homosc. This improvement can be attributed to the flexibility of heterogeneous pricing schemes, which allow for type-specific tolls.

From Figure \ref{fig: SocCost}, we observe that the Price of Anarchy (PoA) -- which is the ratio of the social cost of equilibrium congestion levels induced under no tolls with that of \opt -- is $1.04$ for the Bay area transportation network. This is likely due to the high congestion level of the network during the morning rush hour. Indeed, theoretical studies \citep{colini2020selfish, cominetti2021price} have proved that the PoA approaches to $1$ as the total demand of travelers increases. Moreover, empirical studies \citep{youn2008price, o2016mechanisms, monnot2017bad} have also shown that the PoA in the transportation networks of London, Boston, New York city and Singapore are also close to \(1\). Furthermore, from Figure \ref{fig: SocCost}, we find that all congestion pricing schemes \homo, \homosc, \het, $\hetsc$  outperform $\curr$ in terms of the average travel time. Surprisingly, it is also marginally outperformed by \zero. A key reason is that \curr\ imposes the same tolls on all of the bridges which does not result in effective re-distribution of traffic from eastern corridor to western corridor. 
While a reduced toll price or zero toll price may increase the total demand of travelers, but its impact is likely to be not significant due to (1) the high expense of car ownership and parking fee (\cite{Depillis_Lieberman_Chapman_2023} estimates that US average annual car ownership cost is \$12182 in 2023), and (2) the low coverage of public transportation in the Bay Area.

\begin{figure}[htbp]
    \centering
    \includegraphics[width = 0.5\textwidth]{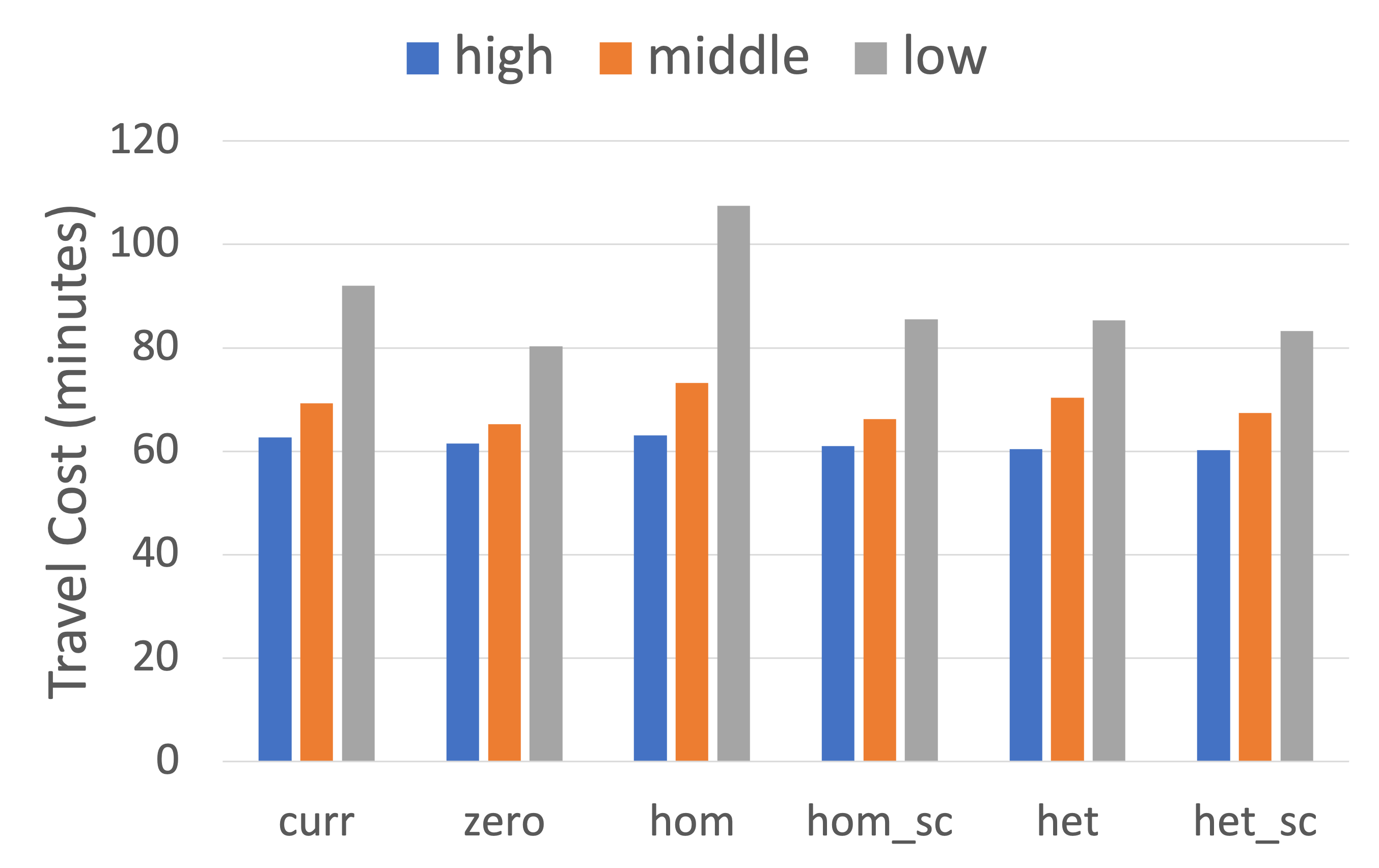}
    \caption{Average travel cost experienced by different types of travelers under different tolling schemes.} 
    \label{fig: ChangeAvgTotCostWrtCurr}
\end{figure}

Figure \ref{fig: ChangeAvgTotCostWrtCurr} illustrates the average travel cost experienced by type of travelers under different pricing schemes. We observe that the difference of average cost across the three traveler types is lower in  \het, \hetsc, \homosc, and \zero, in comparison to \curr.  
Moreover, we observe that for all type of travelers, the average travel cost is lower in \het, \hetsc, \homosc, and \zero, in comparison to \curr.  Furthermore, we note that this observation not only holds in the averaged sense but also in a distributional sense as illustrated in
Table \ref{tab: low_vot_thresh_dist}-\ref{tab: high_dist_threshold}, which presents the proportion of travelers of a particular type experiencing travel costs surpassing a predetermined threshold. We observe that, regardless of the value of threshold and the type of travelers, the proportion of travelers experiencing cost higher than a threshold is higher in \curr\ in comparison to \het, \hetsc, \homosc, and \zero. 
This clearly shows that \curr\ is not preferred by any type of traveler. The pricing scheme \homo\ results in higher travel cost because it cannot differentiate between type of traveler and charges higher tolls to travelers in order to ensure minimum average travel time. 

In homogeneous congestion pricing schemes, regardless of the threshold and the type of traveler, a higher percentage of travelers incur travel costs exceeding a set threshold compared to heterogeneous pricing. This is due to type-specific tolls in heterogeneous schemes resulting in lower tolls for low income travelers. Additionally, pricing schemes with support constraints reduce the percentage of travelers exceeding a threshold. While the differences are marginal between \het\ and \hetsc, such differences are more prominent between \homo\ and \homosc.  

\begin{table}[]
    \centering
    \begin{tabular}{c|c|c|c|c|c|c}
    Travel Cost & \curr &\zero&\hetsc&\homosc&\homo&\het \\
    \hline 
    \hline
    \(\geq\)  60 minutes &69\%  & 56\%  & 64\%  & 67\%  & 76\%  & 66\% \\
  \(\geq\)  90  minutes & 51\%  & 39\%  & 42\%  & 43\%  & 55\% & 42\%\\
  \(\geq\)  120 minutes &32\% & 22\% & 25\% & 27\% & 41\% & 25\%\\
    \(\geq\) 150 minutes &17\% & 10\% & 11\% & 12\% & 26\% & 13\% \\ 
    \hline 
  \end{tabular}
    \caption{low \votTag~travelers }
    \label{tab: low_vot_thresh_dist}
\end{table}
\begin{table}
\centering 
    \begin{tabular}{c|c|c|c|c|c|c}
    Travel Cost & \curr&\zero&\hetsc&\homosc&\homo&\het \\
    \hline 
    \hline
    \(\geq\) 60 minutes &55\%  & 49\%  & 50\%  & 50\%  & 54\%  & 52\% \\
     \(\geq\) 90 minutes & 31\%  & 28\%  & 31\%  & 30\%  & 35\%  & 30\% \\
     \(\geq\) 120 minutes &13\% & 12\%  & 11\%  & 11\%  & 17\%  & 14\% \\
     \(\geq\) 150 minutes &1\%  & 1\%  & 1\%  & 1\%  & 3\%  & 2\% 
\\
\hline
  \end{tabular}
    \caption{middle \votTag~travelers}
    \label{tab: middle_vot_thresh_dist}
\end{table}
\begin{table}[]
    \centering
    \begin{tabular}{c|c|c|c|c|c|c}
    Travel Cost & \curr&\zero&\hetsc&\homosc&\homo&\het \\
    \hline 
    \hline
     \(\geq\)  60 minutes &46\%  & 46\%  & 46\% & 46\%  & 48\%  & 46\% 
\\
     \(\geq\) 90 minutes & 28\%  & 27\%  & 26\%  & 27\%  & 30\%  & 27\% \\
      \(\geq\) 120 minutes &12\%  & 7\%  & 7\% & 8\%  & 10\%  & 7\% \\
     \(\geq\) 150 minutes &1\%  & 0\%  & 0\%  & 0\%  & 1\%  & 1\% \\
    \hline
  \end{tabular}
    \caption{high \votTag~travelers}
    \label{tab: high_dist_threshold}
\end{table}

\begin{figure}[htbp]
    \centering
    \includegraphics[width = 0.6\textwidth]{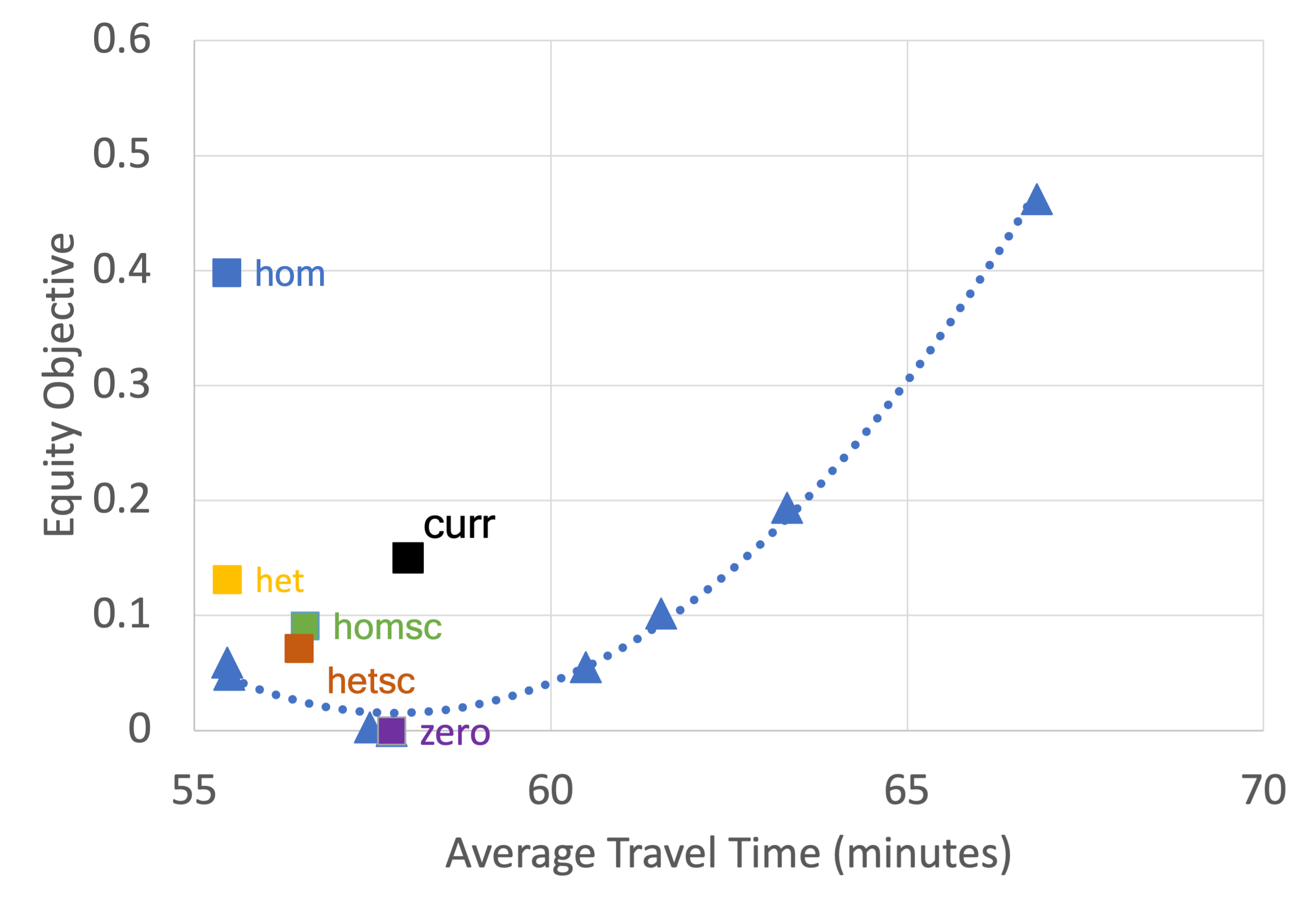}
    \caption{Trade-off between average travel time and equity: The blue triangles represent different pricing schemes, positioned near the Pareto curve (a polynomial best-fit curve through the triangle points), based on computations detailed in the Appendix.} 
    \label{fig: EfficiencyEquityTradeoff}
\end{figure}

Next, in Figure \ref{fig: EfficiencyEquityTradeoff}, we compare different pricing schemes using two metrics: average travel time and the equity metric (as defined in \eqref{eq: planner_obj}-(i)). Our results show that all pricing schemes, except for \homo~, outperform \curr~ on both metrics. Additionally, we present a Pareto front (dotted line) that illustrates the trade-off between minimizing average travel time and reducing inequity. The method used to compute this trade-off curve is detailed in the Appendix.

\subsubsection{Revenue considerations}\label{ssec: RevenueConsideration}
\begin{figure}[htbp]
    \centering
    \includegraphics[width = 0.5\textwidth]{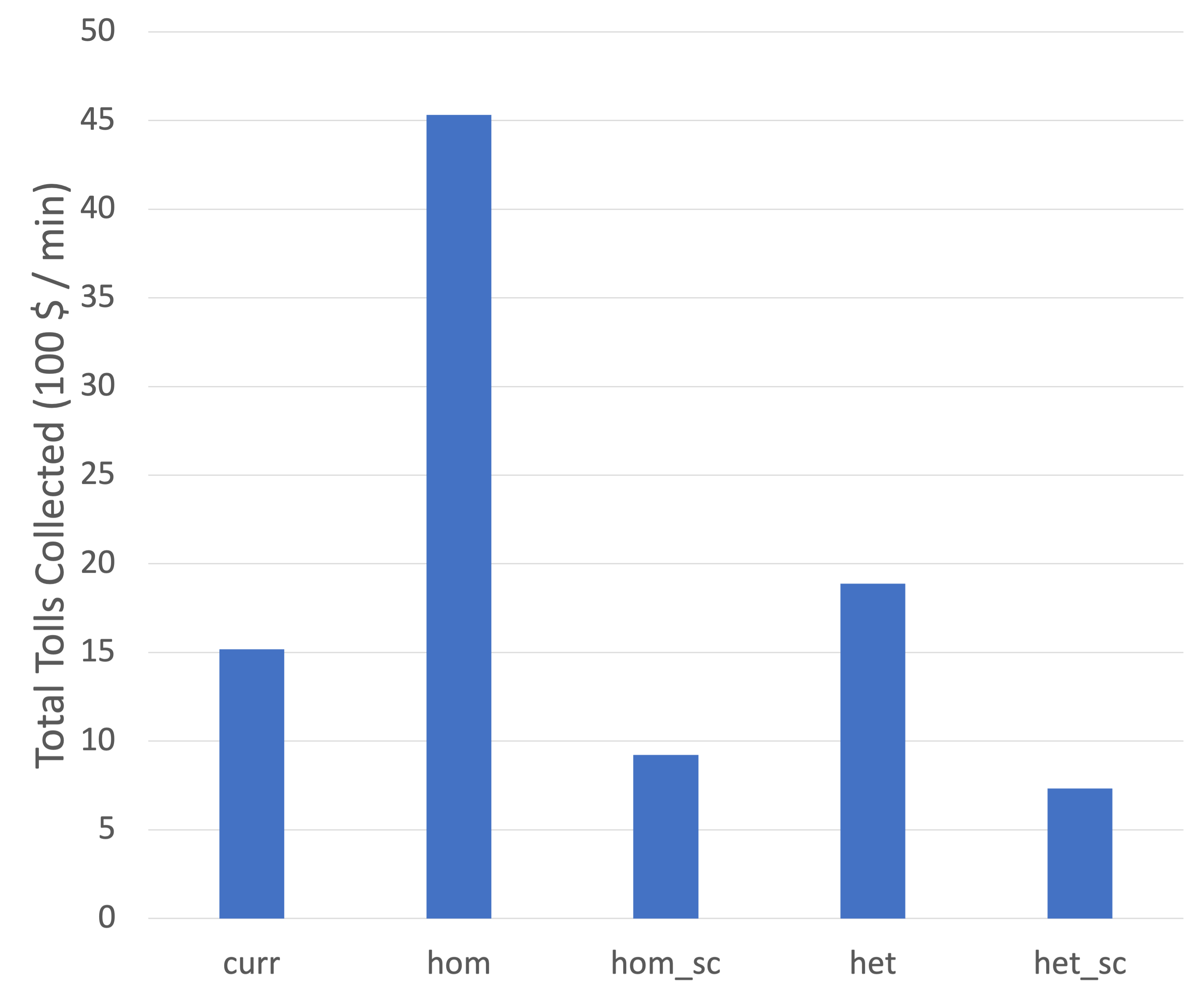}
    \caption{Comparison of total revenue collected for \textsf{current,hom,hom-c,het,het-c}.} 
    \label{fig: TotRevenue}
\end{figure}

Another important aspect of determining the congestion pricing scheme is the revenue it generates, which could be used for maintenance of existing transportation infrastructure, enhancing public transit options, amongst other things. Figure \ref{fig: TotRevenue} presents a comparison of different congestion pricing scheme in terms of total revenue. As per the data released by Metropolitan Transportation Commission (MTC) \footnote{available at \url{https://mtc.ca.gov/about-mtc/authorities/bay-area-toll-authority/historic-toll-paid-vehicle-counts-toll-revenue}} a total toll revenue of $\$633,932,206$ was collected in the Bay Area during the year 2019-2020. Our calibrated model in \curr\ predicts toll revenues on the same order of magnitude but slightly lower than MTC data. The mismatch between our prediction and MTC data is attributed to the fact that \textit{(i)} our analysis only focuses on morning rush hour but MTC data also include tolls collected beyond morning rush hour as well, \textit{(ii)} MTC data also includes tolls on HOV (High Occupancy Vehicle) lanes which are currently not added in our analysis, \textit{(iii)} there is some additional demand incoming from other nearby cities not included in our analysis, and \textit{(iv)} higher tolls are charged to multi-axle vehicles, with tolls charged as high as $\$36$ in 2019.\footnote{refer \url{http://tinyurl.com/MTC-Multi-Axle})}

Notably, \homo\ generates the highest revenue as it applies uniformly higher prices across all edges, irrespective of traveler types, with the goal of achieving a minimum congestion congestion level.
Moreover, the revenue of the other three pricing schemes $\homosc$, $\het$ and $\hetsc$ are comparable to that of $\curr$ with $\het$ being slightly higher and $\homosc$ and $\hetsc$ being slightly lower. 


\section{Conclusion and discussion}\label{sec: Conclusion}
We study the problem of designing congestion pricing schemes which not only minimize the overall congestion but also reduce the disparate impact of congestion pricing schemes on the basis of socioeconomic and geographic diversity of travelers. We present a multi-step linear programming based approach to design four kinds of congestion pricing schemes varying in terms of their implementation depending on whether (a) they can toll travelers on the basis of their willingness-to-pay, and (b) they can toll every edge of the network or only a subset of it. 
The evaluation and comparison of these congestion pricing schemes on the San Francisco Bay Area highway network reveal several significant insights. The proposed schemes outperform the currently implemented scheme in terms of overall congestion reduction and exhibit improvements in equity by providing better travel costs to each type of traveler. The analysis also highlights the revenue generation potential of different pricing schemes.
Furthermore, heterogeneous pricing schemes can yield more equitable distribution of travel cost between different types of travelers, paving the way for future research to explore effective implementation strategies.

There are several interesting directions of future research. First, the implementation question inspires a study of the design of tax rebate programs that facilitate heterogeneous pricing schemes. Second, a more comprehensive empirical understanding of traffic patterns in the Bay Area can be done by incorporating travelers incoming from other cities in the Bay Area. Finally, it would be interesting to account for mode choice between driving and public transit, and more generally to account for elasticity of demand occurring from other choices such as remote work. 

\appendix
\section{Calibration of Latency Functions}
\label{sup:drivingtime}
Here, we present the methodology used to compute the latency function of all freeways in Figure \ref{fig:node_map}. Recall from Section \ref{sec: ModelCalibrationBayArea}, we need to compute the average travel time and average flow on every edge for every day. 
To achieve this goal, we utilize morning rush hour data from the PeMS dataset, spanning from January 2019 to June 2019.
Let's denote the set of all weekdays in this time-frame by \(\mathcal{T}\). For every edge \(e\in E\) and day \(t\in \mathcal{T}\), let's denote the average travel time by $\hat{\ell}_{e}^{t}$ and the average edge flow  by $\hat{w}_{e}^{t}$. In order to estimate these quantities, we use PeMS data during the morning rush hours \(\hourSlot= [6 \text{am}-7 \text{am}, 7\text{am} -8\text{am}, 8\text{am} -9\text{am}, 9\text{am} -10\text{am}, 10 \text{am} -11\text{am}, 11\text{am} -12\text{noon}]\). 
 Let \(\sensor_e\) be the number of sensors fitted on edge \(e\in E\) which provide average hourly flow and average speed information. 

First, we demonstrate how to use the raw data from sensors to compute the average travel time on every edge.  We compute an estimate of the time required to travel the edge \(e\) at hour \(h\) by accumulating the average time required to travel between sensors on that link as follows:
\begin{align}
    \hat{\ell}_{e}^{ht} = \sum_{s=1}^{\sensor_e-1} \frac{\distSensor_e^{s}}{\speedSensor_e^{sht}}, \quad \forall \ e\in E, h \in \mathcal{H}, t\in \mathcal{T}, 
\end{align} 
where \(\distSensor_e^{s}\) is the distance between sensor \(s\) and \(s+1\) on edge \(e\in E\) and \(\speedSensor_e^{sht}\) is the average speed of traffic passing over the sensor \(s\) on edge \(e\) during hour \(h\) on day \(t\). Next, we compute the average hourly flow on an edge as follows: 
\begin{align*}
    \hat{w}_{e}^{ht} = \frac{\sum_{s=1}^{\sensor_e-1} \distSensor_e^s \flowSensor_{e}^{sht}}{\sum_{s=1}^{\sensor_e-1} \distSensor_e^s}, \quad \forall \ e\in E, h \in \mathcal{H}, t\in \mathcal{T},
\end{align*}
where \(\tilde{w}_e^{sht}\) is the hourly average flow of traffic passing over sensor \(s\) on edge \(e\) during hour \(h\) on day \(t\).  
We use the hourly average edge flows $\hat{w}_{e}^{ht}$ and the hourly average travel times $\hat{\ell}_{e}^{ht}$ to compute the average travel time on any edge \(e\in E\) as follows: 
\begin{align*}
    \hat{\ell}_e^t = \frac{\sum_{h\in \hourSlot} \hat{w}_{e}^{ht} \hat{\ell}_{e}^{ht}}{\sum_{h\in \hourSlot} \hat{w}_{e}^{ht}}, \quad e\in E, t\in \mathcal{T}. 
\end{align*}
Similarly, we compute the average of the hourly flows as follows:
\begin{align*}
    \hat{w}_e^t = \frac{1}{|\hourSlot|}\sum_{h\in\hourSlot} \hat{w}_{e}^{ht}, \quad t\in \mathcal{T}, e\in E.
\end{align*}

\section{Calibration of User Demand}
\label{sup:DemandCalibration}
We outline our method for calculating the daily demand of travelers moving between various origin-destination pairs from January 2019 to June 2019. Our approach involves three main steps:

\paragraph{Step 1: Estimating relative demand between nodes using the Safegraph dataset:}
We leverage the Safegraph dataset to obtain the relative demand of travelers traveling between different nodes in the Bay Area. Specifically, the Neighborhood Patterns dataset from Safegraph provides the average daily count of mobile devices moving between different census block groups (CBGs) on workdays for each month. This is then aggregated over the set of nodes after adjusting for sampling bias.

More formally, let's denote the set of CBGs in the Bay Area  by \(\mathcal{C}\). The SafeGraph dataset provides the average daily count of travelers \(N^{cc'}\) traveling from CBG \(c\) to \(c'\). 
However, the SafeGraph dataset exhibits sampling bias\footnote{as referred in {\url{https://colab.research.google.com/drive/1u15afRytJMsizySFqA2EPlXSh3KTmNTQ}}} because different CBGs are sampled at different rates. 
We correct for sampling bias in this data by modifying the counts \(\numCBG^{cc'}\) using the population data provided by the ACS. That is, we compute the corrected count of travelers traveling from CBG \(c\) to \(c'\) as follows 
 \begin{align*}
     \tilde{\numCBG}^{cc'} = \numCBG^{cc'} \frac{R^c}{\sum_{c\in\cbgSet} R^c }\cdot \frac{\sum_{c\in \cbgSet}\sum_{c'\in\cbgSet} \numCBG^{cc'}}{\sum_{c'\in\cbgSet} \numCBG^{cc'}},
 \end{align*}
where \(R^c\) is the number of residents in CBG \(c\) as reported by the ACS dataset. 

\paragraph{Step 2: Calibrating type-specific demands with ACS dataset.}
Given the the adjusted count of travelers we compute the demand of travelers from o-d pair \(k\in K\) by aggregating the demand over set of nodes as follows
\begin{align*}
    \tilde{D}^k = \sum_{c\in k_o}\sum_{c'\in k_d}\tilde{\numCBG}_{\mathcal{C}}^{cc'},
\end{align*}
where $k_o, k_d \in N$ are the origin and destination nodes of the o-d pair \(k\in K\). To obtain the demand in terms of units of flow we compute 
 \begin{align*}
     D^{ik} = \frac{\tilde{D}^k}{\sum_{k'\in K}\tilde{D}^{k'}\mathbbm{1}(k'_o=k_o)} \frac{A^{i k_o}}{|\mathcal{H}|},
 \end{align*}
 where \(A^{i k_o}\) is the total driving population of type \(i\) at node \(k_o\) as given by the ACS dataset and \(|\mathcal{H}|\) is the number of hours in morning rush hours (6 am to 12 noon).

 \paragraph{Step 3: Incorporating daily variability with the PeMS dataset.}
We convert the monthly demand estimates obtained in Step 2 into daily demand data by scaling it proportional to the total daily flow from PeMS dataset. More formally, we compute the average total edge load over all workdays  from January 2019 to June 2019 as follows 
\begin{align}
    \overline{w} = \frac{1}{|\mathcal{T}|} \sum_{t \in \mathcal{T}} \sum_{e \in E} \hat{w}_e^t,
\end{align}
where \(\hat{w}_e^t\) is the average edge load on day \(t\) on edge \(e\), which is obtained in Appendix B using PeMS data. Next, to obtain the daily demand, we scale the monthly demand obtain in Step 2 as follows: 

\begin{align}
    D^{ik}_t = \frac{\sum_{e \in E} \hat{w}_e^t}{\overline{w}}\cdot D^{ik}, \quad \forall\ t\in \mathcal{T}, i\in I, k\in K.
\end{align}


\section{Computing pricing schemes lying on the Pareto curve in Figure \ref{fig: EfficiencyEquityTradeoff}}
Here, we provide a method to compute the Pareto efficient congestion pricing schemes that trade-off between minimizing average travel time and optimizing the equity objective (as in \eqref{eq: planner_obj}).

Before presenting our method to compute Pareto front, we recap the methodology delineated in Section \ref{sec:method}. On a high level, the procedure in Section \ref{sec:method} comprises of two steps: 
\begin{itemize}
    \item \textbf{Step 1:} Characterize the set of tolls that will implement the best possible average travel time \(S(w^\dagger)\)
    \item \textbf{Step 2:} On the set of tolls characterized in Step 1, compute the tolls that optimize the joint equity-welfare objective. 
\end{itemize}
However, the above methodology does not provide a way to compute pricing schemes that trade-off some amount of average travel time in order to improve on equity. Particularly, for any \(S^\ast < S(w^\dagger)\) there is no direct method to characterize the set of pricing schemes that will implement an edge flow vector that would result in average travel time of \(S^\ast\) (as in Step 1 above). 
We provide a new procedure that builds up on the tools presented in Section  \ref{sec:method} to \emph{estimate} this set of tolls which can be use to the tolls that optimize the equity-welfare objective as in Step 2. 

Our approach is stated below: 
\begin{itemize}
    \item Sample \(N\) vectors \(\{\gamma^i\}_{i\in [N]}\subset \mathbb{R}^{|E|}\) such that for every \(i\in [N]\), \(\gamma^i\sim \textsf{Unif}([0,1]^{|E|})\).
    \item For each \(i\in [N]\), solve the following weighted average time minimization problem 
    \begin{align*}
        \min_{w\in W}\sum_{e\in E}\gamma^i_e w_e\ell_e(w_e), 
    \end{align*}
    where \(W\) is the set of all feasible edge flows given demand \(D\) as highlighted below
    \begin{align*}
        W = \{w\in \mathbb{R}^{|E|}: \exists \ q\in \mathcal{Q}(D) \ \text{s.t.} \ w_e = \sum_{i\in I}f_e^i(q)\}. 
    \end{align*}
    Let's denote \(w^{\dagger \gamma^i}\) to be the optimal value of the above optimization problem. 
    \item For each \(i\in [N]\), use \(w^{\dagger\gamma^i}\) in place of \(w^\dagger\) in the optimization process to compute \(\homo\) and \(\het\) pricing schemes (as highlighted in Section \ref{sec:method}).  
    \item Compute the average travel time and equity objective corresponding to each of \(w^{\dagger\gamma^i}\) and compute the Pareto efficient solutions amongst these \(N\) solutions.  
\end{itemize}

In Figure \ref{fig: EfficiencyEquityTradeoff}), the blue triangles are Pareto efficient solutions obtained by taking \(N= 100\) in the above procedure. 
\begin{remark}
    Note that this procedure only provides an estimate of Pareto front and not the exact Pareto front. This is because by definition \(S^\ast_{\gamma^i} := S(w^{\dagger \gamma^i}) \leq S(w^\dagger)\).  
Following similar analysis as in Proposition \ref{prop: HomHetDualOnly}, we can compute the set of pricing schemes that will implement \(w^{\dagger\gamma^i}\) on the transportation network. Unlike \(S(w^\dagger)\), the set of edge flow vectors that result in the average travel time of \(S^\ast_{\gamma^i}\) is not unique, we cannot characterize the entire set of tolls that could result in average travel time \(S^\ast_{\gamma^i}\). Thus, our procedure relies on taking large values of \(N\) so that we can get better estimate of this set.  
s
Extending our approach to derive better estimates of Pareto front is an interesting direction of future research that is bound to help planner in making important design decisions about congestion pricing. 
\end{remark}

\section{Proofs for Section 4}\label{app: ProofsSec}
\textit{Proof of Proposition 1.}
\begin{itemize}
\item[(1)]
To establish this result, we first show that for any given set of tolls \(\tolls\), the optimization problem \eqref{eq: PotentialCongestionGame} is a convex optimization problem. Next, using KKT conditions for optimality we show that the optimal solution to \eqref{eq: PotentialCongestionGame} satisfy the requirements of Nash equilibrium posited in Definition \ref{def: NashEq}. 

To show that the \eqref{eq: PotentialCongestionGame} is a convex optimization problem, we note that the constraint set is convex as it is a product simplex which is a convex set. Next, we show that the objective function is convex. Since the objective is differentiable, it is sufficient to show that 
\begin{align}\label{eq: StrictConvexitySufficientCond}
    \sum_{i\in \votGroup}\sum_{\od\in\odPairs} \sum_{r\in \paths^\od} \left(\frac{\partial \Phi(q,p)}{\partial q_r^{ik}}-\frac{\partial \Phi(\tilde{q},p)}{\partial q_r^{ik}}\right)\left( q_r^{ik} - \tilde{q}_r^{ik} \right) \geq  0 \quad \forall \ q, \tilde{q} \in \mathcal{Q}, q\neq \tilde{q}. 
\end{align}
To see this, we note that 
\begin{align*}
  \frac{\partial \Phi(q,p)}{\partial q_r^{ik}} &= \sum_{e\in \edges} \latency_e(w_e(q)) \frac{\partial w_e(q)} {\partial  q_r^{ik}}   + \sum_{\type\in\votGroup}\sum_{e\in\edges}\frac{(\tolls_{\edge}^{\type}+\gasPrice_e)}{\vot^i}\frac{\partial w_e^i(q)} {\partial  q_r^{ik}} \\ 
  &= \sum_{e\in \edges} \latency_e(w_e(q)) \mathbbm{1}(e\in r)  + \sum_{e\in\edges}\frac{(\tolls_{\edge}^{\type}+\gasPrice_e)}{\vot^i}\mathbbm{1}(e\in r) = c_r^i(q,p).
\end{align*}
    Consequently, for any \(q,\tilde{q}\in \mathcal{Q}\) such that \(q\neq\tilde{q}\) it holds that 
\begin{align*}
     &\sum_{i\in \votGroup}\sum_{\od\in\odPairs} \sum_{r\in \paths^\od} \left(\frac{\partial \Phi(q,p)}{\partial q_r^{ik}}-\frac{\partial \Phi(\tilde{q},p)}{\partial q_r^{ik}}\right)\left( q_r^{ik} - \tilde{q}_r^{ik} \right) \\ 
     &=\sum_{i\in \votGroup}\sum_{\od\in\odPairs} \sum_{r\in \paths^\od} \left(\sum_{e\in \edges} (\latency_e(w_e(q))  -\latency_e(w_e(\tilde{q}))) \mathbbm{1}(e\in r) \right)\left( q_r^{ik} - \tilde{q}_r^{ik} \right),\\
     &=\sum_{e\in \edges} (\latency_e(w_e(q))  -\latency_e(w_e(\tilde{q}))) \sum_{i\in \votGroup}\sum_{\od\in\odPairs} \sum_{r\in \paths^\od} \mathbbm{1}(e\in r) \left( q_r^{ik} - \tilde{q}_r^{ik} \right),\\
     &=\sum_{e\in \edges} (\latency_e(w_e(q))  -\latency_e(w_e(\tilde{q}))) \left( w_e(q) - w_e(\tilde{q}) \right) 
     \geq 0.
\end{align*}
where the last inequality follows because \(\ell_e\) is strictly increasing function. Thus, we have established the \eqref{eq: PotentialCongestionGame} is convex optimization problem. 

Next, we analyze the KKT conditions associated with \eqref{eq: PotentialCongestionGame}. Define the Lagrangian 
\begin{align*}
    \mathcal{L}(q,\lambda,\mu;p) = \Phi(q,p) + \sum_{i\in\votGroup}\sum_{\od\in\odPairs}\lambda^{ik}(D^{ik}-\sum_{r\in\paths^k}q_r^{ik}) -\sum_{i\in\votGroup}\sum_{k\in\odPairs}\sum_{r\in\paths^k}\mu_r^{ik}q_r^{ik}.
\end{align*}
Since \eqref{eq: PotentialCongestionGame} is a convex optimization problem and the strong form of Slater's conditions hold as the feasible set is a product-simplex, we obtain the following first-order necessary and sufficient condition of optimality: 
\begin{align}
    &\frac{\partial \mathcal{L}(q^\ast(p),\lambda^\ast,\mu^\ast;p)}{\partial q_r^{ik}} = 0 \quad \forall \ i\in \votGroup, k\in \odPairs, r\in\paths^k, \tag{C1}\label{eq: C1}\\ 
    &\sum_{r\in\paths^k}q_r^{*ik}(p) = D^{ik}\quad \forall \ i\in \votGroup, k\in \odPairs,\tag{C2}\label{eq: C2}\\
    &\mu_r^{*ik}q_r^{*ik}(p) = 0 \quad \forall \ i\in \votGroup, k\in \odPairs, r\in\paths^k, \tag{C3}\label{eq: C3}\\ 
    &\mu_r^{*ik}\geq 0, q_r^{*ik}(p)\geq 0\quad \forall \ i\in \votGroup, k\in \odPairs, r\in\paths^k.\tag{C4}\label{eq: C4} 
\end{align}
Note that \eqref{eq: C1} can be equivalently written as 
\begin{align*}
    0&=\frac{\partial \mathcal{L}(q^\ast(p),\lambda^\ast,\mu^\ast;p)}{\partial q_r^{ik}}=\frac{\partial \Phi(q^\ast(p),p)}{\partial q_r^{ik}} - \lambda^{ik} - \mu_r^{ik} = c_r^i(q^\ast(p),p) - \lambda^{ik} - \mu_r^{ik}
\end{align*}
Additionally, using \eqref{eq: C4} we obtain that \(c_r^i(q^\ast(p),p) \geq \lambda^{ik} \), for every \(i\in I, k\in K, r\in \paths^k\). Furthermore, from 
\eqref{eq: C3} we obtain that if for some \(i\in I, k\in K, r\in \paths^k\), \(q_r^{\ast ik}> 0\) then \(
    c_r^i(q^\ast(p),p) = \lambda^{ik}.
\) 
This is precisely the conditions stated in Definition \ref{def: NashEq}. 
\item[(2)] 
Using the first-order necessary conditions for constrained optimality, we observe that,
\begin{align}\label{eq: FOCSet1}
     \sum_{i\in \votGroup}\sum_{\od\in\odPairs}\sum_{r\in\paths^k}\frac{\partial S(q^\dagger)}{\partial q_r^{ik}}\left(\tilde{q}_r^{ik}-q_r^{\dagger,ik}\right)\geq  0 \quad \ \forall \ \tilde{q} \in \mathcal{Q}.
\end{align}
Similarly, it holds that 
\begin{align}\label{eq: FOCSet2}
     \sum_{i\in \votGroup}\sum_{\od\in\odPairs}\sum_{r\in\paths^k}\frac{\partial S(\bar{q}^\dagger)}{\partial q_r^{ik}}\left(\tilde{q}_r^{ik}-\bar{q}_r^{\dagger,ik}\right)\geq  0 \quad \ \forall \ \tilde{q} \in \mathcal{Q}.
\end{align}
Selecting \(\tilde{q} = \bar{q}^\dagger\) in \eqref{eq: FOCSet1}, and selecting \(\tilde{q}= q^\dagger\) in \eqref{eq: FOCSet2} and substracting the resulting inequality we obtain 
\begin{align}\label{eq: FOC Constraints}
    \sum_{i\in \votGroup}\sum_{\od\in\odPairs}\sum_{r\in\paths^k}\left(\frac{\partial S(q^\dagger)}{\partial q_r^{ik}}-\frac{\partial S(\bar{q}^\dagger)}{\partial q_r^{ik}}\right) \left({q}_r^{\dagger, ik}-\bar{q}_r^{\dagger,ik}\right) \leq 0.
\end{align} 
Suppose there exists \(q^\dagger, \bar{q}^\dagger \in Q^\dagger\) such that there exists \(e\in E\) such that \(w_e(q^\dagger)\neq w_e(\bar{q}^\dagger)\). Then we will show that \eqref{eq: FOC Constraints} is violated. 

Note that for any \(q\in \mathcal{Q}\),
\begin{align*}
    \frac{\partial S(q)}{\partial q_r^{ik}} &= \sum_{e\in E}\frac{\partial w_e(q)}{\partial q_r^{ik}}\ell_e(w_e(q)) + \sum_{e\in E} w_e(q) \nabla\ell_e(w_e(q))\frac{\partial w_e(q)}{\partial q_r^{ik}} \\ 
    &= \sum_{e\in E}\mathbbm{1}(e\in r)\ell_e(w_e(q)) + \sum_{e\in E} w_e(q) \nabla\ell_e(w_e(q))\mathbbm{1}(e\in r). 
\end{align*}
Using this, we compute the left-hand side of \eqref{eq: FOC Constraints}, 
\begin{align*}
    &\sum_{i\in \votGroup}\sum_{\od\in\odPairs}\sum_{r\in\paths^k}\left(\frac{\partial S(q^\dagger)}{\partial q_r^{ik}} - \frac{\partial S(\bar{q}^\dagger)}{\partial q_r^{ik}}  \right)\left({q}_r^{\dagger, ik}-\bar{q}_r^{\dagger,ik}\right)\\ 
    =& \sum_{i\in \votGroup}\sum_{\od\in\odPairs}\sum_{r\in\paths^k}\sum_{e\in E}\mathbbm{1}(e\in r)(\ell_e(w_e(q^\dagger))-\ell_e(w_e(\bar{q}^\dagger)))\left({q}_r^{\dagger, ik}-\bar{q}_r^{\dagger,ik}\right)\\ &\quad + \sum_{i\in \votGroup}\sum_{\od\in\odPairs}\sum_{r\in\paths^k}\sum_{e\in E}(w_e(q^\dagger) \nabla\ell_e(w_e(q^\dagger))-w_e(\bar{q}^\dagger) \nabla\ell_e(w_e(\bar{q}^\dagger)))\mathbbm{1}(e\in r)  \left({q}_r^{\dagger, ik}-\bar{q}_r^{\dagger,ik}\right)
    \\ 
    =& \sum_{e\in E}(\ell_e(w_e(q^\dagger))-\ell_e(w_e(\bar{q}^\dagger)))\sum_{i\in \votGroup}\sum_{\od\in\odPairs}\sum_{r\in\paths^k}\mathbbm{1}(e\in r)\left({q}_r^{\dagger, ik}-\bar{q}_r^{\dagger,ik}\right)\\ &\quad + \sum_{e\in E}(w_e(q^\dagger) \nabla\ell_e(w_e(q^\dagger))-w_e(\bar{q}^\dagger) \nabla\ell_e(w_e(\bar{q}^\dagger)))\sum_{i\in \votGroup}\sum_{\od\in\odPairs}\sum_{r\in\paths^k}\mathbbm{1}(e\in r)  \left({q}_r^{\dagger, ik}-\bar{q}_r^{\dagger,ik}\right)\\ 
    =& \sum_{e\in E}(\ell_e(w_e(q^\dagger))-\ell_e(w_e(\bar{q}^\dagger)))\left(w_e(q^\dagger)-w_e(\bar{q}^\dagger)\right) \\ &\quad + \sum_{e\in E}(w_e(q^\dagger) \nabla\ell_e(w_e(q^\dagger))-w_e(\bar{q}^\dagger) \nabla\ell_e(w_e(\bar{q}^\dagger)))\left(w_e(q^\dagger)-w_e(\bar{q}^\dagger)\right) 
\end{align*}
Note that 
\(
    \sum_{e\in E}(\ell_e(w_e(q^\dagger))-\ell_e(w_e(\bar{q}^\dagger)))\left(w_e(q^\dagger)-w_e(\bar{q}^\dagger)\right) \geq 0,
\)
due to the monotonicity of latency function. Moreover, note that 
\begin{align*}
    &\sum_{e\in E}(w_e(q^\dagger) \nabla\ell_e(w_e(q^\dagger))-w_e(\bar{q}^\dagger) \nabla\ell_e(w_e(\bar{q}^\dagger)))\left(w_e(q^\dagger)-w_e(\bar{q}^\dagger)\right)\\
    =&\sum_{e\in E}(w_e(q^\dagger) \nabla\ell_e(w_e(q^\dagger))-w_e(\bar{q}^\dagger) \nabla\ell_e(w_e(q^\dagger))+w_e(\bar{q}^\dagger)  \nabla\ell_e(w_e(q^\dagger))-w_e(\bar{q}^\dagger) \nabla\ell_e(w_e(\bar{q}^\dagger)))\cdot \\ &\quad \quad \quad 
    \cdot \left(w_e(q^\dagger)-w_e(\bar{q}^\dagger)\right) \\ 
    =& \sum_{e\in E}\nabla\ell_e(w_e(q^\dagger))(w_e(q^\dagger)-w_e(\bar{q}^\dagger))^2  + \sum_{e\in E}w_e(\bar{q}^\dagger)(\nabla\ell_e(w_e(q^\dagger))-\nabla\ell_e(w_e(\bar{q}^\dagger)))\left(w_e(q^\dagger)-w_e(\bar{q}^\dagger)\right).
\end{align*}
Note that \(\sum_{e\in E}\nabla\ell_e(w_e(q^\dagger))(w_e(q^\dagger)-w_e(\bar{q}^\dagger))^2> 0\) due to the hypothesis that there exists at least one edge where \(w_e(q^\dagger)\neq w_e(\bar{q}^\dagger)\) and the fact that the latency function is strictly increasing. Moreover  \(\sum_{e\in E}w_e(\bar{q}^\dagger)(\nabla\ell_e(w_e(q^\dagger))-\nabla\ell_e(w_e(\bar{q}^\dagger)))\left(w_e(q^\dagger)-w_e(\bar{q}^\dagger)\right)\geq 0\) as \(\ell_e(\cdot)\) is assumed to be convex. Thus, we obtain 
\begin{align*}
    \sum_{i\in \votGroup}\sum_{\od\in\odPairs}\sum_{r\in\paths^k}\left(\frac{\partial S(q^\dagger)}{\partial q_r^{ik}} - \frac{\partial S(\bar{q}^\dagger)}{\partial q_r^{ik}}  \right)\left({q}_r^{\dagger, ik}-\bar{q}_r^{\dagger,ik}\right) > 0,
\end{align*}
which contradicts \eqref{eq: FOC Constraints}. 
\end{itemize}

\hfill $\square$

\textit{Proof of Proposition \ref{prop: HomHetDualOnly}.}

(1) 
First, we prove that given any optimal solution  \((p^\dagger, z^\dagger)\) of \eqref{eq: HomToll}, $p^{\dagger}$ induces the socially optimal edge flow vector $w^{\dagger}$. Consider any optimal solution of \eqref{eq: HomDual}, denoted as \(q^\dagger\). From strong duality theory, we know that \((p^\dagger, z^\dagger, q^\dagger)\) must satisfy complementary slackness conditions associated with the constraints in \eqref{eq: HomToll} and \eqref{eq: HomDual}. In particular, the complementary slackness condition for \eqref{eq: HomToll}-\eqref{eq: HomDual} indicates that for any \(i\in I\), \(k\in K\), and \(r\in \paths^k\), 
\[q^{\dagger ik}_r>0,  \quad \Rightarrow \quad  z^{\dagger ik} = \theta^i\ell_r(w^\dagger) + \sum_{e\in r}(p_e^\dagger+g_e)=\theta^ic_r^i(q^\dagger,p^\dagger).\]
Additionally, from \eqref{eq: HomToll}, we have for all \(i\in I, k\in K, r'\in\paths^k\),
\begin{align*}
 z^{\dagger ik} \leq \theta^i\ell_{r'}(w^\dagger) + \sum_{e\in r'}(p_e^\dagger+g_e)=\theta^i c_{r'}^i(q^\dagger,p^\dagger).   
\end{align*}
Consequently, 
\begin{align}\label{eq:eq_check}
\forall i \in I, k \in K, r \in \paths^k, \quad q^{\dagger ik}_r>0, \quad \Rightarrow \quad c_r^i(q^\dagger,p^\dagger)\leq c_{r'}^i(q^\dagger, p^\dagger), \quad  \forall r' \in R_{k}.
\end{align}
That is, the flow vector $q^{\dagger}$ only takes routes with the minimum cost given the socially optimal edge flow vector $w^{\dagger}$. We next prove that $w^{\dagger}$ is indeed induced by $q^{\dagger}$, i.e. constraint \eqref{eq: HomDual_a} is tight with the optimal solution. 

For notational brevity, we denote $\hat{w}_e = \sum_{i\in I}\sum_{k\in K}\sum_{r\in \{\paths^k| r \ni e\}}q_r^{\dagger ik}$ as the edge flow induced by $q^{\dagger}$. Suppose for the sake of contradiction that for some non-empty subset of edges \(E^\dagger \subseteq E\),
\begin{align*}
    &\forall \ e\in E^\dagger, \quad \hat{w}_e= \sum_{i\in I}\sum_{k\in K}\sum_{r\in \{\paths^k| r \ni e\}}q_r^{\dagger ik} <  w_e^\dagger, \\ 
    &\forall \ e\in E \backslash E^\dagger, \quad  \hat{w}_e= \sum_{i\in I}\sum_{k\in K}\sum_{r\in \{\paths^k| r \ni e\}}q_r^{\dagger ik} =  w_e^\dagger.
\end{align*}
Then, 
\begin{align*}
    &\sum_{e\in E} \hat{w}_e \ell_e(\hat{w}_e ) =   \sum_{e\in E^\dagger} \hat{w}_e\ell_e(\hat{w}_e ) + \sum_{e\in E\backslash E^\dagger} \hat{w}_e \ell_e(\hat{w}_e ) <  \sum_{e\in E^\dagger} w_e^\dagger\ell_e(w_e^\dagger) + 
    \sum_{e\in E\backslash E^\dagger} w_e^\dagger\ell_e(w_e^\dagger) = \sum_{e\in E}w_e^\dagger \ell_e(w_e^\dagger),
\end{align*}
where the inequality is due the the fact that \(\ell_e\) is a strictly increasing function. This contradicts with the fact that \(w^\dagger\) minimizes the social cost function. Therefore, we must have \(\hat{w}_e = \sum_{i\in I}\sum_{k\in K}\sum_{r\in \{\paths^k| r \ni e\}}f_r^{\dagger ik} = w_e^\dagger\), for every \(e\in E\). 

Following from the fact that $q^{\dagger}$ satisfies \eqref{eq:eq_check} and induces the socially optimal edge flow vector $w^{\dagger}$, we can conclude that \(w^\dagger\) is an equilibrium edge flow vector induced by the flow vector $q^{\dagger}$ associated under the toll price \(p^\dagger\). Hence, the optimal solution $p^{\dagger}$ of \eqref{eq: HomToll} indeed implements the socially optimal edge flow.  

We now prove the other direction. Suppose that there exists a $\hom$ toll vector \(\tilde{p}\) that induces the socially optimal edge flow \(w^\dagger\) in equilibrium, then there exists \(\tilde{z}\) such that \((\tilde{z}, \tilde{p})\) is an optimal solution to \eqref{eq: HomToll}. We denote \(\tilde{q}\) as a Nash equilibrium strategy distribution given toll \(\tilde{p}\). Then, such \(\tilde{q}\) is a feasible solution of \eqref{eq: HomDual}, and \eqref{eq: HomDual_a} holds with equality. 

Next, we define \(\tilde{z}^{ik} = \min_{r\in \paths^k} \theta^i\tilde{\ell}_r(w^\dagger) + \sum_{e\in r}(\tilde{p}_e + g_e)\). This ensures that 
\begin{align*}
    \tilde{z}^{ik} \leq \theta^i\tilde{\ell}_r(w^\dagger)+\sum_{e\in r}(\tilde{\tolls}_e+\gasPrice_e), \quad \forall k\in K, r\in \paths^k, i\in I. 
\end{align*}
Therefore, \((\tilde{p}, \tilde{z})\) is a feasible solution of the primal problem \eqref{eq: HomToll}. Moreover, we note that \((\tilde{p}, \tilde{z}, \tilde{q})\) satisfies the complementary slackness condition associated with \eqref{eq: HomToll} and \eqref{eq: HomDual}. Thus, \((\tilde{p},\tilde{z})\)  is an optimal solution to \eqref{eq: HomToll}  and \(\tilde{q} 
\) is an optimal solution to \eqref{eq: HomDual}.

(2) The proof of this part follows an analogous procedure as that in part (1). We denote an optimal solution of \eqref{eq: Het} as \((p^\dagger, z^\dagger)\), and an optimal solution of \eqref{eq: HetDual} as \(q^\dagger\). From the complementary slackness condition associated with \eqref{eq: Het}-\eqref{eq: HetDual}, we know that if \(q^{\dagger ik}_r>0\) for some \(i\in I, k\in K, r\in \paths^k\), then \(z^{\dagger ik} = \theta^i\tilde{\ell}_r(w^\dagger) + \sum_{e\in r}(p_e^{\dagger i}+g_e)=\theta^i c_r^i(q^\dagger,p^\dagger)\). Moreover, we know that for every \(i\in I, k\in K, r'\in\paths^k\),
\begin{align*}
 z^{\dagger ik} \leq \theta^i\tilde{\ell}_{r'}(w^\dagger) + \sum_{e\in r'}(p_e^{\dagger i}+g_e)=\theta^i c_{r'}^i(q,p^\dagger),   
\end{align*}
which implies that \(c_r^i(q^\dagger,p^\dagger)\leq c_{r'}^i(q^\dagger, p^\dagger)\), i.e.  \(q^\dagger\) sends flow on routes with the minimum cost associated with the heterogeneous toll \(p^\dagger\)  and the socially optimal edge flow $w^{\dagger}$. Moreover, following the same procedure as that in the $\hom$ case, we can argue that $q^{\dagger}$ induces the socially optimal edge flow $f^{\dagger}$ (i.e. \eqref{subeq: het_1} is tight), otherwise we arrive at a contradiction that $f^{\dagger}$ is not socially optimal. Therefore, we can conclude that $q^\dagger$ is an equilibrium strategy distribution that induces the socially optimal (type-specific) edge flow $f^{\dagger}$ given the $\het$ toll vector $p^\dagger$. 

On the other hand, suppose that there exists a $\het$ toll vector \(\tilde{p}\) that induces the socially optimal edge flow \(w^\dagger\)in equilibrium, then we define \(\tilde{z}^{ik} = \min_{r\in \paths^k} \theta^i \ell_r(w^\dagger) + \sum_{e\in r}(\tilde{p}^i_e + g_e)\) for all \( k\in K\), \(r\in \paths^k\), and \(i\in I\). Analogous to the case with $\hom$ toll, we can argue that  \((\tilde{p},\tilde{z})\) (resp. \(\tilde{q} 
\) ) is a feasible solution of \eqref{eq: Het} (resp. \eqref{eq: HetDual}), and satisfies complementary slackness conditions. Consequently, we know that  \((\tilde{p},\tilde{z})\) (resp. \(\tilde{q} 
\) ) is an optimal solution of \eqref{eq: Het} (resp. \eqref{eq: HetDual}). 
 \hfill $\square$

\bibliography{refs}
\bibliographystyle{apalike}
\end{document}